\documentclass[]{aastex631}
\usepackage{amsmath}

\definecolor{DCW}{rgb}{0,0.7,0}
\definecolor{Hiro}{rgb}{0,0.5,1}

\received{ }
\revised{ }
\accepted{ }

\submitjournal{ApJ}

\shorttitle{Closure Relations of Optical GRBs}

\graphicspath{{./}{figures/}}

\begin{document}

\title{The closure relations in optical Afterglow of Gamma-Ray Bursts}

\correspondingauthor{M. G. Dainotti}
\email{maria.dainotti@nao.ac.jp}

\author[0000-0003-4442-8546]{M. G. Dainotti}
\affiliation{National Astronomical Observatory of Japan, Tokyo, Japan}
\affiliation{Sokendai University, Japan}
\affiliation{Space Science Institute, Boulder, Colorado}

\author[0000-0003-3411-6370]{D. Levine}
\affiliation{Department of Astronomy, University of Maryland, College Park, MD 20742, USA}

\author[0000-0002-0173-6453]{N. Fraija}
\affiliation{Instituto de Astronomia, Universidad Nacional Autónoma de México}

\author{D.Warren}
\affiliation{RIKEN Interdisciplinary Theoretical and Mathematical Sciences Program (iTHEMS), Wakō, Saitama, 351-0198 Japan}

\author{S. Sourav}
\affiliation{Department of Engineering Sciences, Indian Institute of Science Education and Research,Bhopal,Madhya Pradesh, India}

\begin{abstract}
Gamma-ray bursts (GRBs) are extremely high-energy events that can be observed at very high redshift. In addition to $\gamma$ rays, they can emit in X-ray, optical, and sometimes radio wavelengths. Here, following the approach in \citet{Srinivasaragavan2020ApJ, 2021PASJ...73..970D, 2021ApJS..255...13D} and Dainotti et al (2022, submitted), we consider 82 GRBs from \citet{Dainotti2022} that have been observed in optical wavelengths and fitted with a broken power law (BPL). We consider the relations between the spectral and temporal indices (closure relations; CRs) according to the synchrotron forward-shock model evolving in the constant-density interstellar medium (ISM; $k=0$) and the stellar Wind environment ($k=2$) in both slow- and fast-cooling regimes, where the density profile is defined as $n \propto r^{-k}$. We find the $\nu > \rm max\{\nu_{\rm c}, \nu_{\rm m}\}$ regime is most favored, where $\nu_{\rm c}$ and $\nu_{\rm m}$ are the cooling and characteristic frequencies, respectively. Finally, we test the 2D Dainotti correlation between the rest-frame end time of the plateau and the luminosity at that time on GRBs that fulfill the most-favored CRs. When we compare the intrinsic scatter $\sigma_{int}$ of those 2D correlations to the scatter presented in \cite{Dainotti2020ApJb,Dainotti2022}, we see the scatters of our correlations generally agree with the previous values within $1 \sigma$, both before and after correction for selection bias. This new information has helped us to pinpoint subsamples of GRBs with features that could drive the GRB emission mechanism, and eventually allow for GRBs to be used as standard candles.
\end{abstract}

\section{Introduction} \label{sec:intro}
Gamma-ray bursts (GRBs) are very high energy bursts that can emit in $\gamma$-rays, X-rays, optical, and radio wavelengths. They are of significant interest as their high luminosities allow them to be observed at great distances, including redshifts up to $z=10$ \citep{cucchiara11}. This has sparked interest in using GRBs as standard candles, for a better method of probing cosmological distances.

GRBs undergo two phases - the ``prompt" episode, which is the initial explosion seen in $\gamma$-rays, X-rays, and sometimes optical wavelengths where the phase of fast rise and exponential decay is visible, and the longer-lasting ``afterglow" emission, which can be observed in X-ray, optical, and sometimes radio wavelengths. Beginning in 2004, the Neil Gehrels Swift Observatory (hereafter \emph{Swift}) has observed many GRB afterglows at a wide range of redshifts, luminosities, and durations. To date, \emph{Swift} has observed 1543 GRBs, which includes 1301 UV and optical observations supplemented with UVOT (UltraViolet and Optical Telescope). This makes optical afterglow observations the second-most frequently recorded, following 1347 X-ray observations with the X-ray Telescope (XRT) \footnote{\url{https://swift.gsfc.nasa.gov/archive/grb_table/stats/}}. The multiwavelength observations from \emph{Swift} have been essential to GRB population studies, as they allow for comparison among wavelengths and classes that give vital insight into the physics of GRBs. 

GRB afterglows are thought to be caused by the external forward shock (FS), which is when the relativistic outflow from the progenitor impacts the external medium \citep{Paczynski1993, Katz+97,meszaros97, 1999A&AS..138..537S, 2000ApJ...532..286K, Kumar2010MNRAS}. The FS is an integral part of the widely-accepted standard fireball model \citep{1995ApJ...455L.143S, 1996ApJ...473..204S, 1999A&AS..138..537S, 2000ApJ...532..286K, 2002ApJ...568..820G,zhang06}, which describes the emission of the GRB in both the prompt and afterglow phases. The fireball model can be tested with closure relations, or relations between the spectral and temporal indices of a GRB. These typically aim to determine whether observations of a GRB agree with expectations of two environments - the constant-density interstellar medium (ISM), which is generally assumed to be the environment for short GRBs (SGRB), or the stellar-wind environment, which is ejected during the collapse of massive stars \citep{1994MNRAS.269L..41M, DaigneMochkovitch98, Panaitescu+00,2000ApJ...536..195C}.

Studies of closure relations have previously been conducted in high-energy, X-rays, optical, and radio. Extensive sets of CRs have been presented in studies such as \citet{2004IJMPA..19.2385Z, Racusin+09, Gao2013, Liang2017, 2019ApJ...883..134T}. In high-energy, \citet{2019ApJ...883..134T} studied a sample of 59 GRBs observed by Fermi-LAT (Large Area Telescope) in both a constant-density ISM and Wind environment. They found that their sample is well-described by the standard fireball model, and their most preferred environments are those that involve Slow Cooling (SC-only) or both slow- and fast cooling in the ISM environment.
Dainotti et al (2022, submitted) also considered high-energy emission in a sample of 86 GRBs observed by Fermi-LAT and similarly found that their sample can be well-described by the standard fireball model with a preference for SC-only or slow- and fast-cooling environments. 

In X-rays, a study by \citet{Racusin+09} considered CRs both with and without energy injection for a sample of 230 X-ray afterglows. They found that the CRs are satisfied by a majority of their sample and the standard fireball model describes their sample well. They also found a preference for the ISM environment over the Wind environment. A more recent study by \citet{Srinivasaragavan2020ApJ} considered a sample of 455 X-ray GRB afterglows after the plateau phase on CRs without energy injection and again found that their sample fit the standard fireball model and prefers SC-only regimes, with no strong preference between an ISM or Wind environment. \citet{2021PASJ...73..970D} considered the same sample, but within the plateau region on CRs with energy injection, and found a particular preference for an SC, Wind environment, with the majority of their sample satisfying at least one CR. Studies of synchrotron self-Compton emission in X-rays reveal that CRs in X-rays can mimic the traditional power-law decay phase if thermal electrons are present \citep{2022ApJ...924...40W}. 

In optical, a prior study by \citet{oates2012} considered 48 GRBs observed by \emph{Swift} and found that their sample follows the basic standard fireball model according to the CRs. They tested three CRs, one in a Wind environment, one in an ISM environment, and one independent of an external environment, and found low rates of satisfaction for each relation. Other studies include both X-ray and optical wavelengths, such as \citet{wang15}, which considered a sample of 85 GRBs observed with \emph{Swift} and found 45 GRBs that were well-described by the standard fireball model, with an additional 37 GRBs described at least in part by the fireball model. \citet{2017ApJ...844...92F} considered numerical models of X-ray and optical light curves in an ISM environment in both fast- and slow-cooling regimes, applying their model to GRB 130427A. They found discrepancies in the decay indices for X-ray and optical wavelengths, suggesting that a more complicated model is needed to accurately describe the GRB afterglow. \cite{2022A&A...662A.126J} studied the optical afterglow of long GRB 190919B and found the behavior of the afterglow consistent with an SC, ISM environment. Some studies have considered both optical and X-ray afterglows of single GRBs - \citet{2011A&A...526A.154A} studied GRB 050502B to decipher possible correlations between optical and X-ray data and found both the optical and X-ray afterglow satisfy SC relations for both ISM and Wind, while \citet{2009MNRAS.400...90S} studied GRB 080721 and found the afterglow inconsistent with expectations of the standard fireball model. 

Though it is the least frequently observed afterglow, recent studies have considered CRs in radio wavelengths. \citet{2021ApJ...911...14K} examined a sample of 21 radio light curves (LCs) and found their sample does not fulfill the CRs well, indicating their sample is largely incompatible with the standard fireball model, as the behavior of the radio LCs does not agree with the behavior in the corresponding X-ray and optical LCs. \citet{2021MNRAS.504.5685M} studied the afterglow of GRB 190114C in multiple wavelengths, including X-ray, optical, and radio, and found the X-ray and radio LCs are incompatible with the standard model. 
For instance, \cite{2019ApJ...879L..26F,2019ApJ...883..162F} found that synchrotron self-Compton (SSC) emissions from reverse and forward shock regions are required to describe successfully the afterglow observations of GRB 190114C.
GRB afterglows at low energy are especially susceptible to the presence of thermal electrons in the particle distribution
\citep{2022ApJ...924...40W,2017ApJ...845..150R}, the tension between observations and the traditional model for afterglows may point to a revised understanding of how astrophysical shocks energize the particles that they sweep up. On the other hand, \cite{2022arXiv220611490F} considered the Second Gamma-ray Burst LAT Catalog (2FLGC) and SSC afterglow model evolving in a constant and stratified medium, and when the progenitor injects continuous energy into the blastwave. They found that the CRs of the SSC model can satisfy a significant fraction of bursts in the 2FLGC that cannot be interpreted in the synchrotron afterglow scenario.

Previous studies have shown that \emph{Swift} X-ray LCs have complex features beyond a simple power-law decay \citep{Tagliaferri2005,Nousek2006,OBrien2006,Zhang2006,2007ApJ...666.1002Z, zhang07, zhang07a, zhang07b,sakamoto07,Zhang2019, 2019ApJ...885...29F}. A plateau feature or a flattening of the LC immediately following the prompt emission and preceding the afterglow decay, has been found in all wavelengths of the GRB afterglow, including X-ray \citep{OBrien2006,Zhang2006,Nousek2006, sakamoto07,Evans2009}, optical \citep{Dainotti2020ApJb, Dainotti2022}, and radio \citep{2022ApJ...925...15L} wavelengths. These plateaus typically last from on the order of $10^2$ seconds to $10^5$ seconds, and are thought to result from continuous energy injection from a central engine \citep{dai98,rees98,sari2000,zhang2001,Zhang2006,liang2007} - either the fall-back accretion of matter onto a black hole  \citep{Kumar2008,Cannizzo2009,cannizzo2011,Beniamini2017,Metzger2018} or the spin-down luminosity from a newborn magnetar \citep{zhang2001,troja07,Toma2007,dallosso2011,rowlinson2013,rowlinson14,rea15,BeniaminiandMochkovitch2017,Metzger2018,Stratta2018,Fraija2020}. 


This plateau has been used to develop a correlation between the rest-frame end time of the plateau in X-rays and its correspondent luminosity at that time, called the 2D Dainotti correlation in X-rays \citep{Dainotti2008, dainotti2010, Dainotti11b, Dainotti2013b, delvecchio16, Dainotti2017a, Li2018b}. This correlation has been extended to optical \citep{Dainotti2020ApJb, Dainotti2022} and
radio afterglows \citep{2022ApJ...925...15L}. It has been proposed as a possible method of standardizing the varied sample of GRBs to use it as a cosmological tool  \citep{cardone09, cardone10, Dainotti2013a, 2014ApJ...783..126P, 2022arXiv220607479D, Dainotti2022Gal, 2022MNRAS.514.1828D, 2022arXiv220408710C, 2022MNRAS.512..439C}. \citet{2016ApJ...825L..20D, Dainotti2017b, Dainotti2020a, Dainotti2022} analyze an extension of this two-dimensional correlation to the three-dimensional fundamental plane both in X-rays and optical, which describes a correlation between the rest-frame end time of the plateau, the luminosity at that time, and the peak prompt luminosity.


Generally, interpreting GRB afterglows in optical wavelengths presents a more complicated problem than other wavelengths. In addition to the forward shock, the afterglow could involve a possible reverse shock (material that propagates back into the GRB shell; \citet{1995ApJ...455L.143S}), or an association with a supernova — both of which readily emit at optical wavelengths. This makes testing CRs on optical GRBs difficult, as the CRs test the afterglow resulting specifically from the FS and may not be able to accurately describe emissions resulting from these other physical phenomena.
Nevertheless, in this study, we aim to determine whether a sample of optical GRB afterglows can be accurately described by the standard fireball model through a set of CRs. We also aim to determine if GRBs classified as agreeing with a particular environment can follow the two-dimensional luminosity-time correlation and whether those correlations have a reduced scatter from the optical two-dimensional correlation found in the previous literature.

This paper is organized as follows: in \S\ref{sec:sample} we discuss the sample used for this analysis, in \S\ref{sec:methods} we discuss our methodology, both in terms of the CRs (\S\ref{sec:methodCR}) and the luminosity-time correlation (\S\ref{sec:methodcorr}). In \S\ref{sec:CRs}, we present the results of the CR analysis in optical, and in \S\ref{sec:compare}, we compare the results in optical to results obtained in other wavelengths. In \S\ref{sec:correlation}, we present the results of the correlation analysis with preferred CRs. Finally, in \S\ref{sec:conclusion}, we discuss our results and present our conclusions. 

\section{Data Sample and Fitting} \label{sec:sample}
We take our sample of 82 GRBs from a larger sample of 99 GRBs in \citet{Dainotti2022}. The GRBs have been observed in optical wavelengths by satellites such as Swift and the Ultraviolet/Optical Telescope (UVOT), as well as ground-based detectors such as the Subaru Telescope, the MITSuME, the Re-ionization and Transients InfraRed telescope (RATIR), and the Gamma-Ray Burst Optical/Near-IR Detector (GROND). The sample of 99 GRBs has been fitted with a broken power law (BPL): 
\begin{equation}
    F(t) = 
    \begin{cases} 
      F_a (\frac{t}{T_a})^{-\alpha_1} & t<T_a \\
      F_a (\frac{t}{T_a})^{-\alpha_2} & t \geq T_a\,,
   \end{cases}
   \end{equation}
where $T_{\rm a}$ is the break time, $F_{\rm a}$ is the flux at $T_{\rm a}$, and $\alpha_1$ and $\alpha_2$ are the temporal indices for times before and after $T_{\rm a}$, respectively. We only select GRBs that present a plateau, which we define as those LCs for which $|\alpha_1| < 0.5$. The full sample of the optical GRBs is presented in Table \ref{table:sample}.

We take the temporal indices to be $\alpha_2$ from the BPL fitting, as $\alpha_2$ describes the light curve after the end of the plateau emission, which corresponds to phase III of the LC as in \citet{zhang06}. We consider the segment of the LC that occurs after the plateau but before the jet break. We gather $\beta$ from the literature following the convention $F_v \propto t^{-\alpha}v^{-\beta}$, where $\alpha$ is the temporal index and $\beta$ is the spectral index. Due to the difficulty of gathering a large sample of spectral indices at the specific time $T_a$, we take the $\beta$ parameter from the literature and assume a constant value throughout the LC for all GRBs in our sample. For 30 GRBs in which no optical spectral index could be found in the literature, \citet{Dainotti2022} extrapolates the spectral index from the photon index $\Gamma$ in X-ray (given in the \emph{Swift} XRT repository, \citet{Evans2009}) by assuming $\beta = \Gamma-1$. However, it cannot always be guaranteed that the X-ray and optical LCs fall in the same region of the spectral energy distribution (SED), therefore, for these GRBs, we compare the temporal index found in the optical fitting to the corresponding temporal index found in the X-ray fitting \citep{Srinivasaragavan2020ApJ}. If the temporal indices agree within $1 \sigma$, we can assume that the X-ray and optical LCs fall within the same region (private communication by Bing Zhang) and take the extrapolated spectral index, allowing us to retain 13 GRBs in our sample. Otherwise, we remove the GRBs from our sample. We thus remove 17 GRBs from the sample of 99, giving us a total sample of 82 GRBs. 

\begin{figure}
\includegraphics[width=6cm, height=5cm]{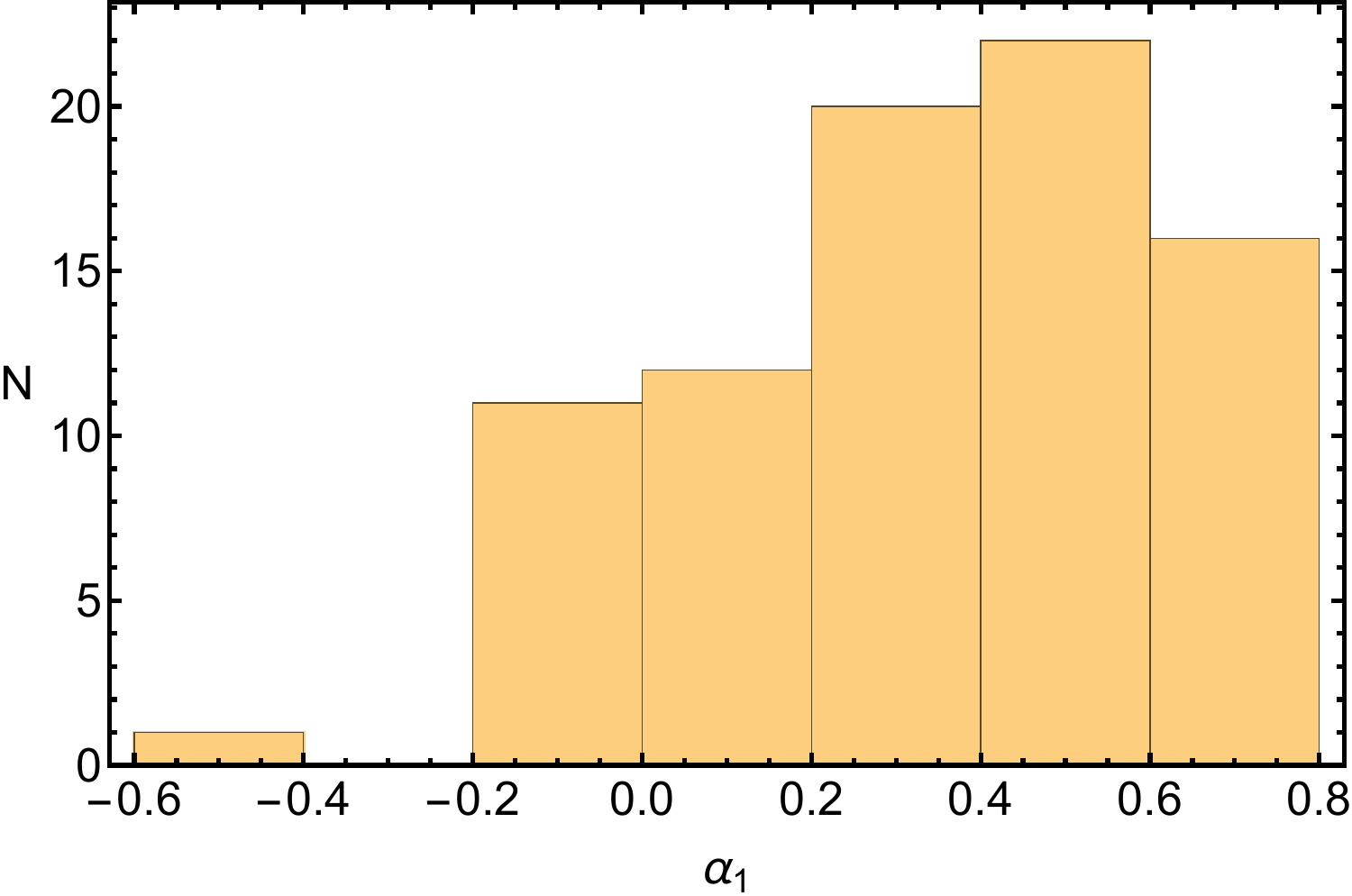}
\includegraphics[width=6cm, height=5cm]{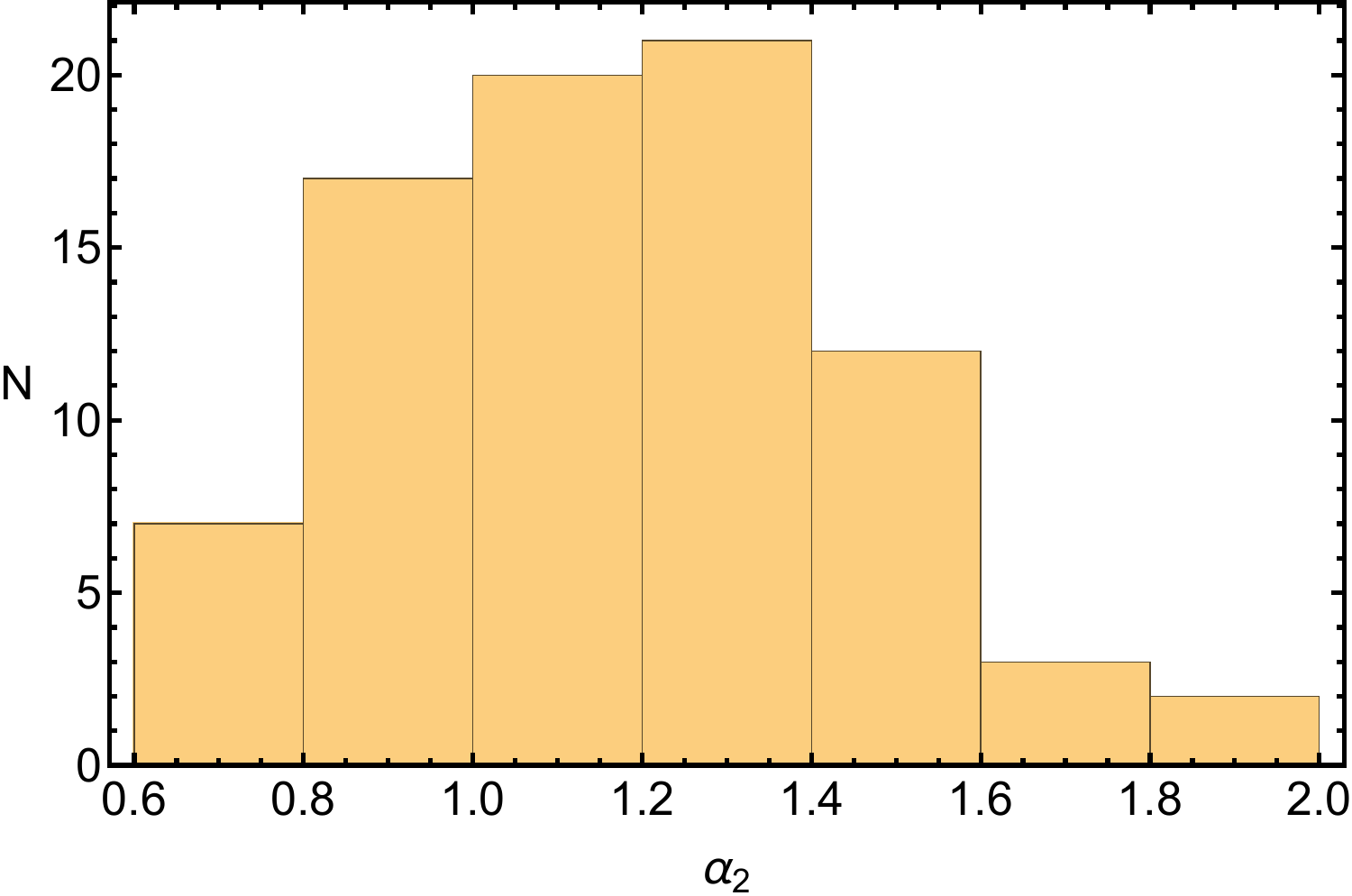}
\includegraphics[width=6cm,height=5cm]{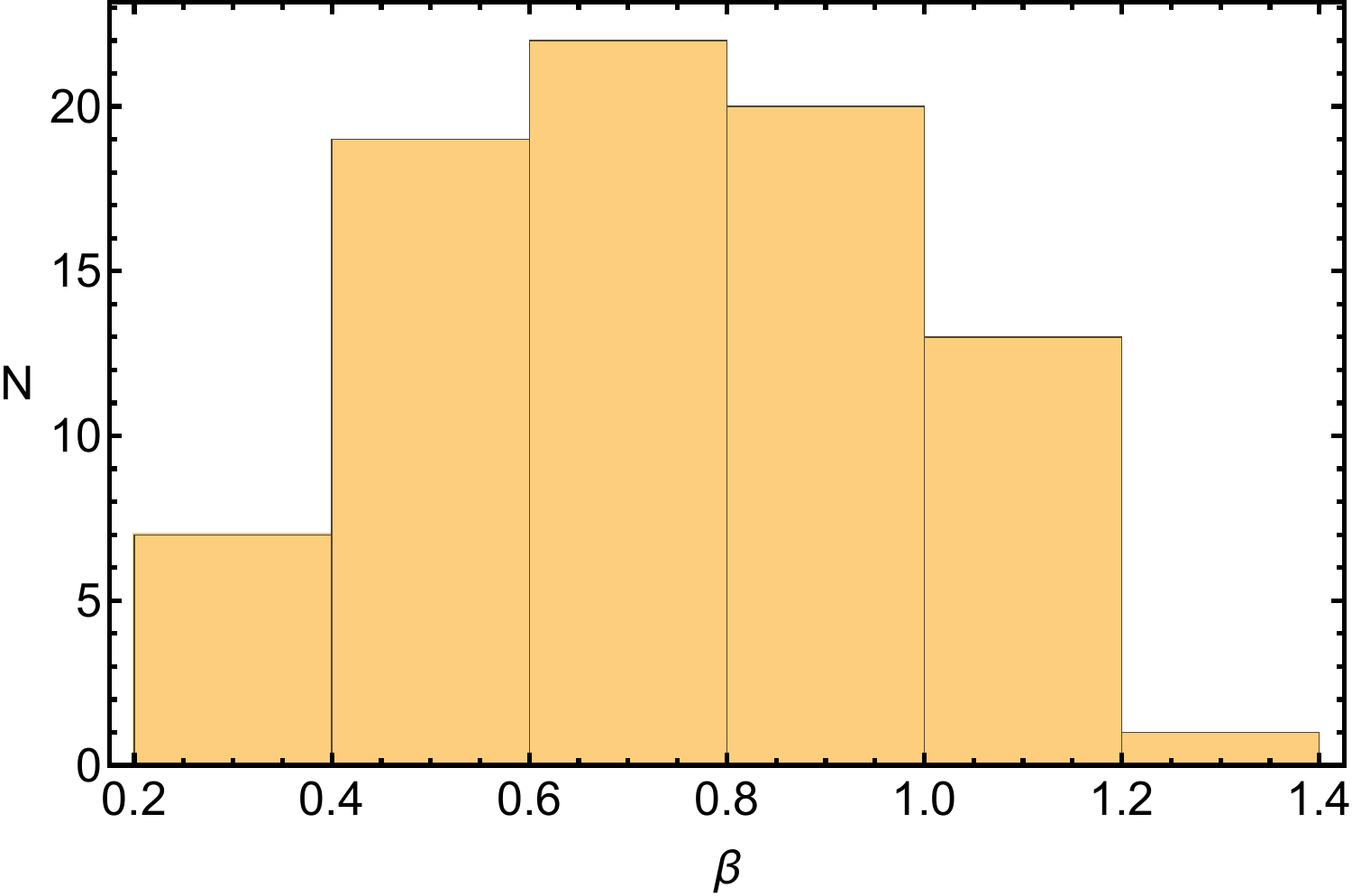}
\caption{Distribution of $\alpha_1$ (left), $\alpha_2$ (middle), $\beta$ (right) for the 82 GRBs used in this study.}.
\label{fig:distIndices}
\end{figure}

\begin{figure}
\includegraphics[width=9cm, height=7cm]{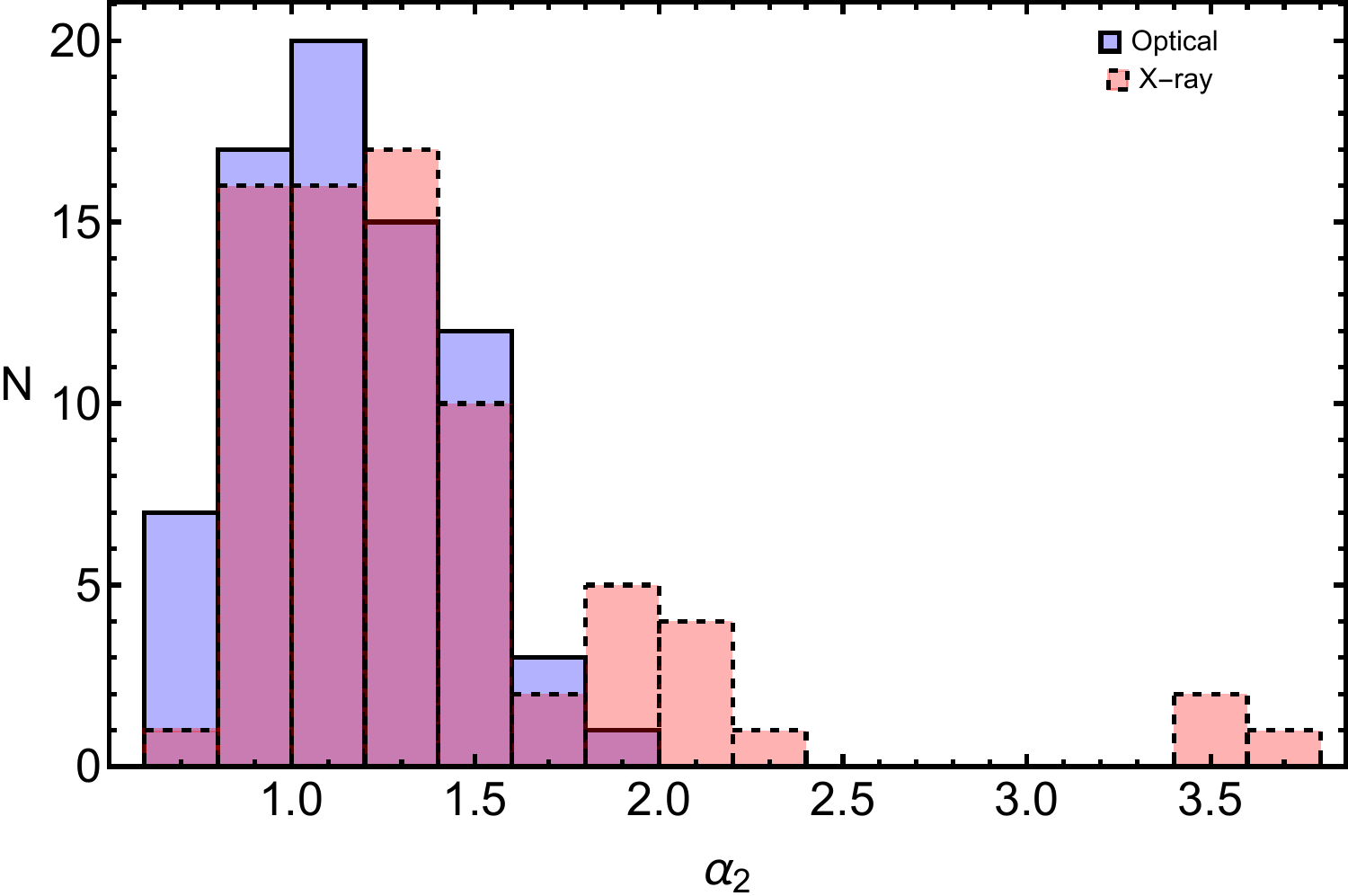}
\includegraphics[width=9cm, height=7cm]{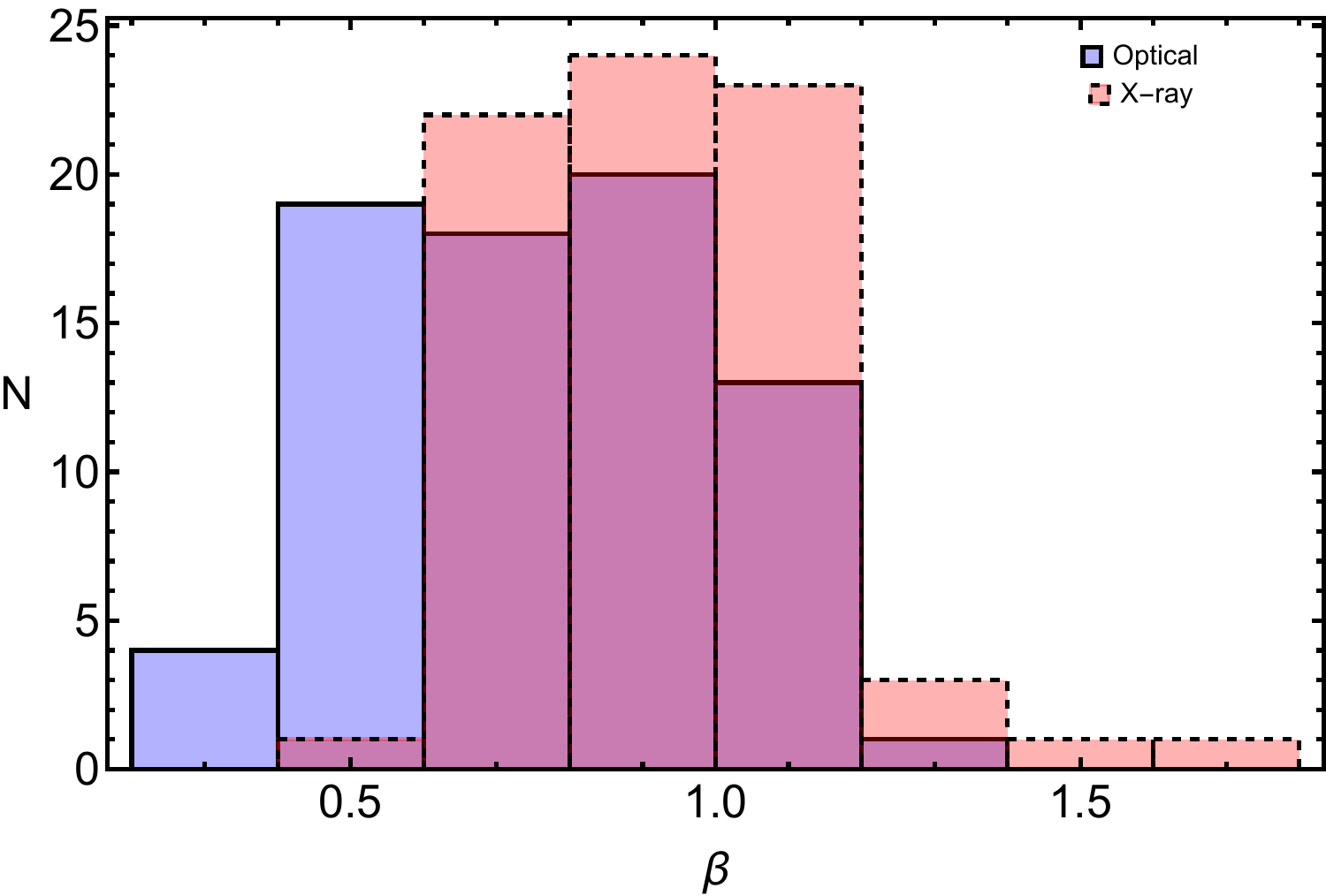}
\caption{Distribution of $\alpha_2$ (left) and $\beta$ (right) for sample of optical GRBs and the corresponding values in X-ray. GRBs observed pre-Swift and GRBs not found in the Swift XRT catalog are not included in the plot.}
\label{fig:XOdist}
\end{figure}

For an example of an optical LC fitted with a BPL in our sample, please refer to \citet{Dainotti2022}. We show the distributions of the $\alpha_1$, $\alpha_2$, and $\beta$ parameters in our sample in Fig. \ref{fig:distIndices}. For comparison, we also show the corresponding values of $\alpha_2$ and $\beta$, as measured between the end of the plateau and the end of the LC, in X-ray wavelengths in Fig. \ref{fig:XOdist} for the sample of optical GRBs. \footnote{In Fig \ref{fig:XOdist}, we do not include GRBs observed pre-Swift (before 2005) or those not found in the XRT catalog, as the spectral index could not be computed.} 

\startlongtable
\begin{deluxetable*}{lCCcccC}
\tablecaption{Sample of 82 GRBs used in this analysis, fit with a BPL. The first two columns show the $\alpha_2$, where $\alpha_2$ is the distribution of the temporal index after the end of the plateau, and $\beta$ values with their errors, the third and fourth columns show the reference for the LC and the spectral indices, respectively. The sixth column shows the class of the GRB - either long (LGRBs) or ultra-long (UL); short (SGRBs), including intrinsically short (IS) and short with extended emission (SEE); X-ray flashes (XRF), X-ray rich GRBs (XRR), or GRBs with supernova (SNe) associations, classified as A, B, C, D, or E according to \citet{hjorth03}. The seventh column shows the rest-frame time of break $T^*_a$ found from the BPL fitting, where the * denotes the rest frame.}\label{tab:sample}
\tablewidth{0.5pt}
\tabletypesize{\scriptsize}
\tablehead{\colhead{GRB} & \colhead{${\alpha} \pm \delta_{\alpha}$}  & \colhead{$\beta \pm \delta_{\beta}$} & \colhead{author} & \colhead{$\beta_{ref}$} & \colhead{class} & \colhead{$log{T^*_{a}}$ (s)}}
\startdata
GRB010222A	&	1.25	\pm	0.08	&	0.76	\pm	0.22   &   [1], [2] &[3]& L&4.24\\
GRB021004A	&	1.38	\pm	0.03	&	0.67	\pm	0.14   &   [1]     &[4]& L&4.87\\
GRB030328A	&	1.30	\pm	0.03	&	0.36	\pm	0.45   &   [1]     &[4]& L&3.97\\
GRB040924A	&	1.30	\pm	0.01	&	0.63	\pm	0.48   &   [1]     &[4]&SN-C&3.23\\ 
GRB041006A	&	1.24	\pm	0.01	&	0.36	\pm	0.27   &   [5]     &[3]&SN-C&3.84\\
GRB050319A	&	0.81	\pm	0.02	&	0.74	\pm	0.42   &   [6]     &[4]&XRR&3.81\\
GRB050401A	&	0.88	\pm	0.04	&	0.39	\pm	0.05   &   [1], [7] &[1]&L&3.55\\ 
GRB050416A	&	0.83	\pm	0.05	&	0.92	\pm	0.30   &   [1]     &[3]&XRF-D-IS-SN&3.93\\
GRB050502A	&	1.89	\pm	0.21	&	0.76	\pm	0.16   &   [8]     &[4]&L&3.03\\
GRB050525A	&	1.38	\pm	0.02	&	0.52	\pm	0.08   &   [3]     &[4]&SN-B-XRR&3.01\\
GRB050730A	&	1.46	\pm	0.04	&	0.52	\pm	0.05   &   [3]     &[3]&L&3.64\\
GRB050801A	&	1.18	\pm	0.00	&	0.69	\pm	0.34   &   [3]     &[4]&XRR&2.26\\
GRB050820A	&	1.07	\pm	0.01	&	0.72	\pm	0.03   &   [3], [6]  &[4]&L&3.90\\
GRB050824A	&	0.73	\pm	0.01	&	0.45	\pm	0.18   &   [3]     &[4]&XRF-SN-E&3.38\\
GRB050922C	&	1.28	\pm	0.00	&	0.56	\pm	0.01   &   [3], [6], [9] &[4]&IS&3.26\\
GRB051111A	&	1.27	\pm	0.08	&	0.76	\pm	0.07   &   [5]     & [1] & L & 2.36\\
GRB060124A	&	0.87	\pm	0.00	&	0.73	\pm	0.08   &   [6]     &[10]&XRR&3.11\\
GRB060206A	&	1.24	\pm	0.01	&	0.77	\pm	0.01   &   [6]     &[4]&XRR-IS&3.59\\
GRB060512A	&	0.77	\pm	0.02	&	0.68	\pm	0.05   &   [3]     &[1]&XRF&3.14\\
GRB060526A	&	1.05	\pm	0.01	&	0.65	\pm	0.06   &   [3]     &[4]&XRR&3.56\\
GRB060607A	&	1.21	\pm	0.03	&	0.72	\pm	0.27   &   [3]     &[4]&L&2.92\\
GRB060708A	&	0.81	\pm	0.03	&	0.88	\pm	0.05   &   [11]    &[12]&L&2.69\\
GRB060714A	&	0.99	\pm	0.04	&	0.44	\pm	0.04   &   [5]     &[1]&XRR&3.64\\
GRB060729A	&	1.20	\pm	0.03	&	0.67	\pm	0.07   &   [6]     &[4]&XRR-SN-E&4.88\\
GRB060904B	&	1.16	\pm	0.01	&	1.11	\pm	0.10   &   [3]     &[4]&XRR-SN-C&3.69\\
GRB060927A	&	1.40	\pm	0.09	&	0.86	\pm	0.03   &   [3]     &[1]&XRR&2.42\\
GRB061126A	&	1.20	\pm	0.06	&	0.82	\pm	0.09   &   [1], [3]    &[3]&L&3.24\\
GRB070110A	&	1.00	\pm	0.01	&	1.00	\pm	0.14   &   [3]     &[1]&XRR&3.91\\
GRB070411A  &   1.34    \pm 0.09    &   1.09    \pm 0.11   &   [6]     &[13]&XRR&2.72\\
GRB070810A	&	1.53	\pm	0.15	&	1.01	\pm	0.08   &   [3]     &[14]&XRR&3.26\\
GRB071003A	&	1.70	\pm	0.01	&	0.35	\pm	0.11   &   [3]     &[3]&L&5.09\\
GRB071031A	&	0.75	\pm	0.00	&	0.34	\pm	0.30   &   [3]     &[4]&XRF&2.97\\
GRB071112C	&	0.92	\pm	0.00	&	0.44	\pm	0.11   &   [4]     &[4]&SN-C&2.29\\
GRB080310A	&	1.08	\pm	0.00	&	0.42	\pm	0.12   &   [6]     &[4]&XRR&2.99\\
GRB080319C	&	1.18	\pm	0.03	&	0.98	\pm	0.42   &   [3]     &[4]&L&2.76\\
GRB080330A	&	1.11	\pm	0.00	&	0.42	\pm	0.15   &   [3]     &[4]&XRF&3.12\\
GRB080413A	&	1.47	\pm	0.12	&	0.52	\pm	0.37   &   [5]     &[4]&XRR&1.35\\
GRB080603A	&	0.95	\pm	0.02	&	0.85	\pm	0.31   &   [6]     &[4]&L&3.37\\
GRB080603B	&	1.34	\pm	0.01	&	0.62	\pm	0.06   &   [4]     &[4]&L&3.63\\
GRB080605A	&	0.69	\pm	0.01	&	0.57	\pm	0.35   &   [4]     &[4]&L&3.10\\
GRB080710A	&	1.52	\pm	0.00	&	0.80	\pm	0.09   &   [3]     &[4]&L&3.74\\
GRB080913A	&	0.90	\pm	0.13	&	1.16	\pm	0.17   &   [3]     &[4]&IS&4.65\\
GRB081203A	&	1.53	\pm	0.41	&	1.08	\pm	0.02   &   [15]    &[11]&L&3.34\\
GRB090313A	&	0.93	\pm	0.03	&	1.00	\pm	0.10   &   [6]     &[4]&XRR&2.84\\
GRB090423A	&	0.91	\pm	0.09	&	0.45	\pm	0.13   &   [3]     &[4]&XRR-IS&3.73\\
GRB090426A	&	1.17	\pm	0.02	&	0.76	\pm	0.14   &   [5]     &[1]&S-IS-XRR&2.01\\
GRB090510A	&	0.96	\pm	0.14	&	0.85	\pm	0.05   &   [1]     &[16]&S&3.34\\
GRB090927A	&	1.26	\pm	0.16	&	0.41	\pm	0.16   &   [4]     &[4]&SEE-IS&3.93\\
GRB091020A	&	1.01	\pm	0.02	&	1.06	\pm	0.09   &   [4]     &[4]&L&2.15\\
GRB091024A	&	1.53	\pm	0.03	&	0.64	\pm	0.29   &   [4]     &[4]&UL&3.64\\
GRB091127A	&	1.14	\pm	0.00	&	0.43	\pm	0.10   &   [5]     &[1]&XRR-SN-B&4.24\\
GRB100418A	&	1.11	\pm	0.14	&	0.98	\pm	0.09   &   [5]     &[11]&XRF-SN-D&4.65\\
GRB100513A	&	1.45	\pm	1.23	&	1.20	\pm	0.10   &   [17]    &[13]&L&3.56\\
GRB100621A	&	1.58	\pm	0.02	&	0.78	\pm	0.09   &   [4]     &[4]&XRR&3.99\\
GRB100906A	&	1.08	\pm	0.11	&	0.84	\pm	0.22   &   [1]     &[4]&XRR&2.16\\
GRB101219B	&	0.96	\pm	0.00	&	0.58	\pm	0.07   &   [4]     &[4]&XRR-SN-B&3.67\\
GRB110503A	&	1.49	\pm	0.40	&	0.80	\pm	0.04   &   [18]    &[13]&L&4.29\\
GRB110715A	&	1.47	\pm	0.08	&	0.90	\pm	0.22   &   [4]     &[4]&XRR&4.92\\
GRB110726A	&	0.69	\pm	0.02	&	0.50	\pm	0.19   &   [4]     &[4]&L&2.91\\
GRB120119A	&	1.31	\pm	0.03	&	0.89	\pm	0.01   &   [5]     &[11]&L&3.00\\
GRB120404A	&	1.49	\pm	0.02	&	1.05	\pm	0.09   &   [1]     &[11]&XRR&3.37\\
GRB120711A	&	1.19	\pm	0.10	&	0.53	\pm	0.02   &   [16]    &[16]&L&4.01\\
GRB120815A	&	0.60	\pm	0.01	&	0.78	\pm	0.01   &   [5]     &[11]&L&2.73\\
GRB120907A	&	0.87	\pm	0.16	&	0.70	\pm	0.06   &   [19]    &[13]&L&2.54\\
GRB130606A	&	1.23	\pm	0.03	&	0.83	\pm	0.12   &   [4]     &[4]&XRR&3.41\\
GRB130702A	&	1.36	\pm	0.01	&	0.44	\pm	0.09   &   [4]     &[4]&SN-A-XRF&5.00\\
GRB140423A	&	1.02	\pm	0.02	&	0.54	\pm	0.08   &   [20]    &[4]&L&3.55\\
GRB140430A	&	0.80	\pm	0.02	&	1.05	\pm	0.08   &   [4]     &[4]&XRR&3.42\\
GRB140801A	&	0.86	\pm	0.02	&	0.67	\pm	0.16   &   [21]    &[4]&L&3.83\\
GRB140907A	&	1.00	\pm	0.08	&	1.00	\pm	0.07   &   [22]    &[13]&L&3.85\\
GRB141221A	&	1.14	\pm	0.02	&	0.26	\pm	0.09   &   [4]     &[4]&XRR&2.42\\
GRB160131A	&	1.25    \pm	0.56	&	0.03	\pm	0.04   &   [23]    &[13]&L&3.70\\
GRB160227A  &   1.1     \pm 0.18    &   0.70    \pm 0.04   &   [24]    &[13]&L&3.47\\
GRB160804A  &   1.0     \pm 0.13    &   0.90    \pm 0.07   &   [25]    &[13]&L&4.48\\
GRB161219B  &   0.91    \pm 0.04    &   0.49    \pm 0.01   &   [26]    &[4]&L-SN-C&6.47\\
GRB170714A  &   1.92    \pm 0.69    &   0.90    \pm 0.03   &   [27]    &[13]&UL&4.20\\
GRB180205A  &   1.06    \pm 0.07    &   0.90    \pm 0.09   &   [28]    &[13]&L&4.63\\
GRB180325A	&   1.37    \pm 0.11    &   1.13    \pm 0.10   &   [29]    &[4]&L&3.40\\
GRB191011A  &   0.74    \pm 0.67    &   0.90    \pm 0.09   &   [30]    &[13]&L&2.86\\
GRB201020A  &   1.67    \pm 0.83    &   1.10    \pm 0.11   &   [31]    &[13]&L&3.20\\
GRB970508A	&   1.27    \pm 0.03    &   0.32    \pm 0.15   &   [33], [3]     &[4]&XRF&5.98\\
GRB210210A  &   1.65    \pm 0.18    &   0.81    \pm 0.07   &   [32]    &[13]&L&3.64 \\
\enddata
\tablecomments{Data taken from  \citet{Dainotti2022}, [1] \citet{Tang2012APJ}, [2] \citet{Watanabe2001PASJ}, [3] \citet{kann2010}, [4] Kann et al 2022 in preparation, [5] \citet{Si18}, [6] \citet{zaninoni13}, [7] \citet{kamble09}, [8] \citet{2005GCN..3340....1D}
[9] \citet{2009MNRAS.395..490O}, [10] \citet{misra07}, [11] \citet{Li2015APJ}, [12] Oates et al 2022 in preparation, [13] \citet{Evans2009}, [14] \citet{Schady12}, [15] \citet{2008GCN..8603....1D}, [16]\citet{Shao2018ApJS}, [17] \citet{2010GCN.10751....1E}, [18] \citet{2011GCN.12003....1F}, [19] \citet{2012GCN.13721....1Y},  [20] \citet{2014GCN.16164....1P}, [21] \citet{2014GCN.16666....1K}, [22] \citet{2014GCN.16791....1K}, [23] \citet{2016GCN.18961....1M}, [24] \citet{2016GCN.19103....1Y}, [25] \citet{2016GCN.19761....1M}, [26] \citet{2016GCN.20296....1D}, [27] \citet{2017GCN.21351....1M}, [28] \citet{2018GCN.22396....1E}, [29] \citet{2018GCN.22544....1S}, [30] \citet{2019GCN.26002....1K}, [31] \citet{2020GCN.28711....1K}, [32] \citet{2021GCN.29502....1K}, [33] \citet{kann06}. For GRBs with LC data taken from multiple GCNs, please refer to \citet{Dainotti2022} for the complete list of references.}
\label{table:sample}
\end{deluxetable*}

We note that the distributions of the temporal and spectral indices differ slightly between the optical and X-ray samples. We run the Kolmogorov-Smirnov two-sample test for these distributions, finding a p-value of $p = 0.001$, which indicates that these distributions were not drawn from the same parent sample. However, the KS test has limitations: it is best suited for continuous distributions, it has increased sensitivity toward the center of the distribution and less toward the tails of the distribution, and sensitivity to differences in multiple characteristics between the two distributions (including location, scale and shape). In this case, the KS test is likely to be too restrictive to provide us with critical results. Thus, we have approached the problem in a different way. Usually the distributions of GRB parameters in X-rays and optical are well fitted by Gaussian distributions (Dainotti et al. 2022 in preparation), thus we perform a less restrictive and more general test by fitting the distributions with Gaussians. The optical distribution has a mean, $\mu = 0.75$ and a standard deviation, $\sigma = 0.23$, and the X-ray has $\mu = 0.92$ with $\sigma = 0.20$. Thus, the difference between the X-ray and optical samples is not statistically significant - we find that the distributions are compatible within one $\sigma$.

\section{Methodology} \label{sec:methods}

\subsection{Analyzing CRs} \label{sec:methodCR}

Following a similar procedure to Dainotti et al. (2022, submitted), we test our sample with a set of closure relations (CRs), namely, relations between the temporal indices ($\alpha$) and spectral indices ($\beta$). We consider the constant-density ISM ($k=0$) and stellar wind medium ($k=2$) without injection presented in \citet{2019ApJ...883..134T} in slow-cooling (SC) and fast-cooling (FC) regimes. The CRs are defined in two regimes according to the electron spectral index $p$, with $1 < p < 2$ and $p > 2$. The temporal and spectral indices of each GRB are plotted with their error bars. As we expect the $\alpha$ and $\beta$ errors to be correlated, we represent these with ellipses around the points rather than with rectangular shapes. We define the ``fulfillment" of the relations as the intersection of the GRB with the CR at any point within the $1 \sigma$ ellipses. 

\begin{figure*}
\centering
\includegraphics[height = 9cm]{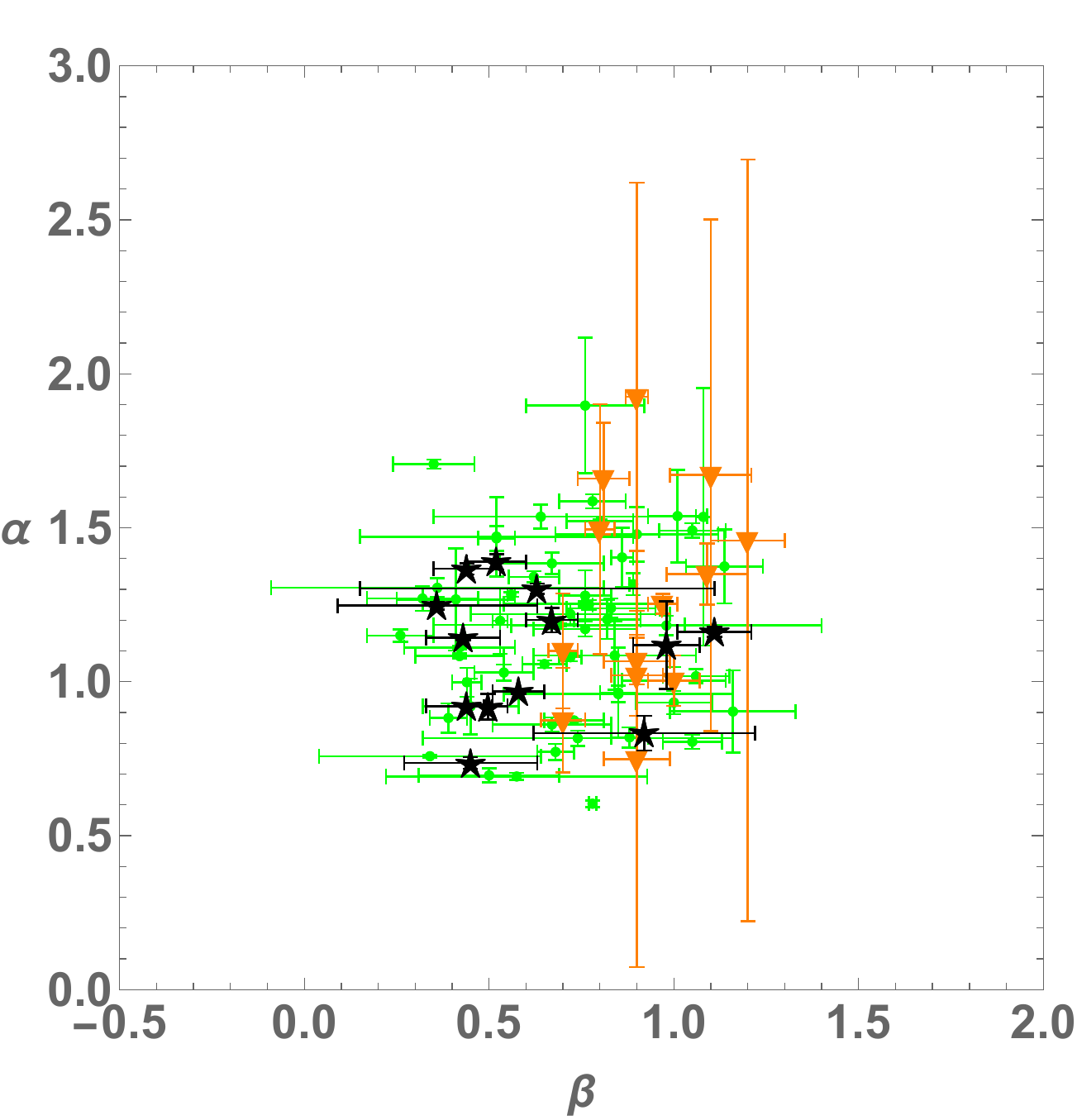}
\caption{Full sample of 82 GRBs considered in this sample. GRBs with $\beta$ extrapolated from X-rays are shown as orange triangles, while GRBs with known SNe associations are shown as black stars.}
\label{fig:sample}
\end{figure*}

Within our sample, we have 13 GRBs that have a spectral index extrapolated from the photon index in X-rays (orange triangles in Fig. \ref{fig:sample}). It is clear that the orange points are clustered at higher values of $\beta$, but we consider them anyway in the analysis since the intent here is to analyze the CRs for the largest possible sample that presents optical plateau emissions. We also have 13 GRBs that have known SNe-Ib/c associations, which have more complex emissions than the FS model. We consider these 13 GRBs-SNe in the total sample as well as separately to determine their degree of fulfillment and if their behavior differs from that of the full sample. We show the full sample in Fig. \ref{fig:sample} classified according to these two considerations - GRBs with an extrapolated spectral index are shown as orange triangles, and GRBs with known SNe associations are shown as black stars. All other GRBs are shown as green circles. We keep this convention throughout the paper. We show the number and rate of the fulfillment of the CRs for the total sample in Table \ref{table:crSummary} and the GRBs-SNe in Table \ref{table:crSNe}. A visual representation of the CRs for all GRBs is shown in Fig. \ref{fig:noinj}. 

\subsection{Computing the 2D luminosity-time correlation}\label{sec:methodcorr}

For the 82 GRBs in our sample, we also compute the rest-frame isotropic luminosity $L_a$ according to: 
\begin{equation}
L_\text{a}= 4 \pi D_L^2(z) \, F_\text{a} \textit{K}
\label{eq:lum}
\end{equation}
where $D_L(z)$ is the luminosity distance assuming a flat $\Lambda$CDM cosmological model with $\Omega_M=0.3$ and $H_0=70$ km s$^{-1}$ Mpc$^{-1}$. $F_a$ is the measured optical energy flux (erg cm$^{-2}$ s$^{-1}$) at time $T_a$, the end of the plateau, and $K$ is the \textit{K}-correction $K=1/(1+z)^{1-\beta}$ \citep{Bloom2001}, where $\beta$ is the optical spectral index.

We use the luminosity to develop the two-dimensional correlation between the rest-frame end time of the plateau $T^*_a$, where the * indicates the rest frame, and the luminosity at that time, $L_a$. The 2D relation is defined similarly to its use in other wavelengths \citep{Dainotti2020a,Dainotti2022}:
\begin{equation}
\log L_\text{a} = c + a \times \log T^{*}_\text{a}  \label{the fundamental plane}
\end{equation}

where $a$ is the slope of the correlation and $c$ is the normalization constant. 

We consider this correlation for the samples of GRBs that fulfill a CR following \citet{Srinivasaragavan2020ApJ,2021PASJ...73..970D} to determine if the intrinsic scatter of the optical correlation can be reduced by GRBs that follow a particular CR. The correlation is analyzed under two conditions, both before and after correction for selection bias and redshift evolution. To apply this correction, we use the Efron \& Petrosian \citep{Efron1992,1995ApJ...449..216E} method to remove the evolution of the data with redshift and recover the intrinsic behavior of the data. This method has been extensively tested in previous GRB studies. \citep{Dainotti2013a,Dainotti2015b, Dainotti15a,Dainotti2017a,Dainotti2017b,Dainotti2017c,Dainotti2020a,2021Galax...9...95D, 2022ApJ...925...15L, Dainotti2022,  2022arXiv220607479D, 2022MNRAS.514.1828D}.

\section{Results} \label{sec:results}

\subsection{Optical closure relations} \label{sec:CRs}

We see that the greatest rates of fulfillment are within the $\nu > \rm max\{\nu_{\rm c}, \nu_{\rm m}\}$ regime, with 19 GRBs fulfilling the CR for $k=0$ and 18 GRBs fulfilling the CR for $k=2$ (Table \ref{table:crSummary}). The second-most preferred regime for the CRs is the SC, $\nu_{\rm m} < \nu < \nu_{\rm c}$ regime, with 2 GRBs fulfilling the CR for $k=0$ and one GRB fulfilling the CR for $k=2$. The least-preferred regime is the FC, $\nu_{\rm c} < \nu < \nu_{\rm m}$, with 0\% fulfillment for both values of $k$. In all CRs tested, we see no strong preference for either the ISM or Wind medium. 

\begin{figure}[ht!]
\centering
\gridline{\fig{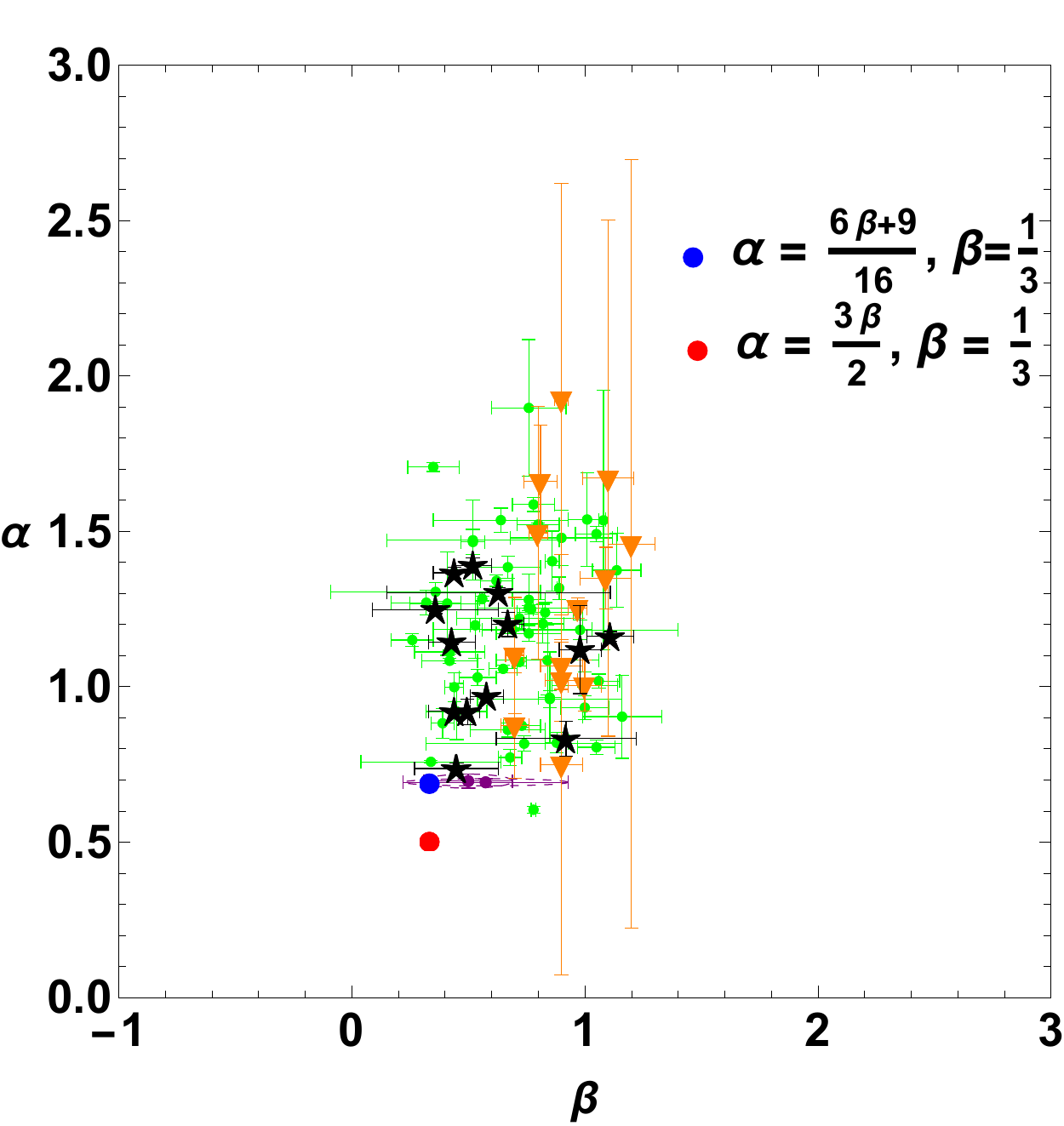}{0.3\textwidth}{k = 0, SC, $v_m < v < v_c$}
\fig{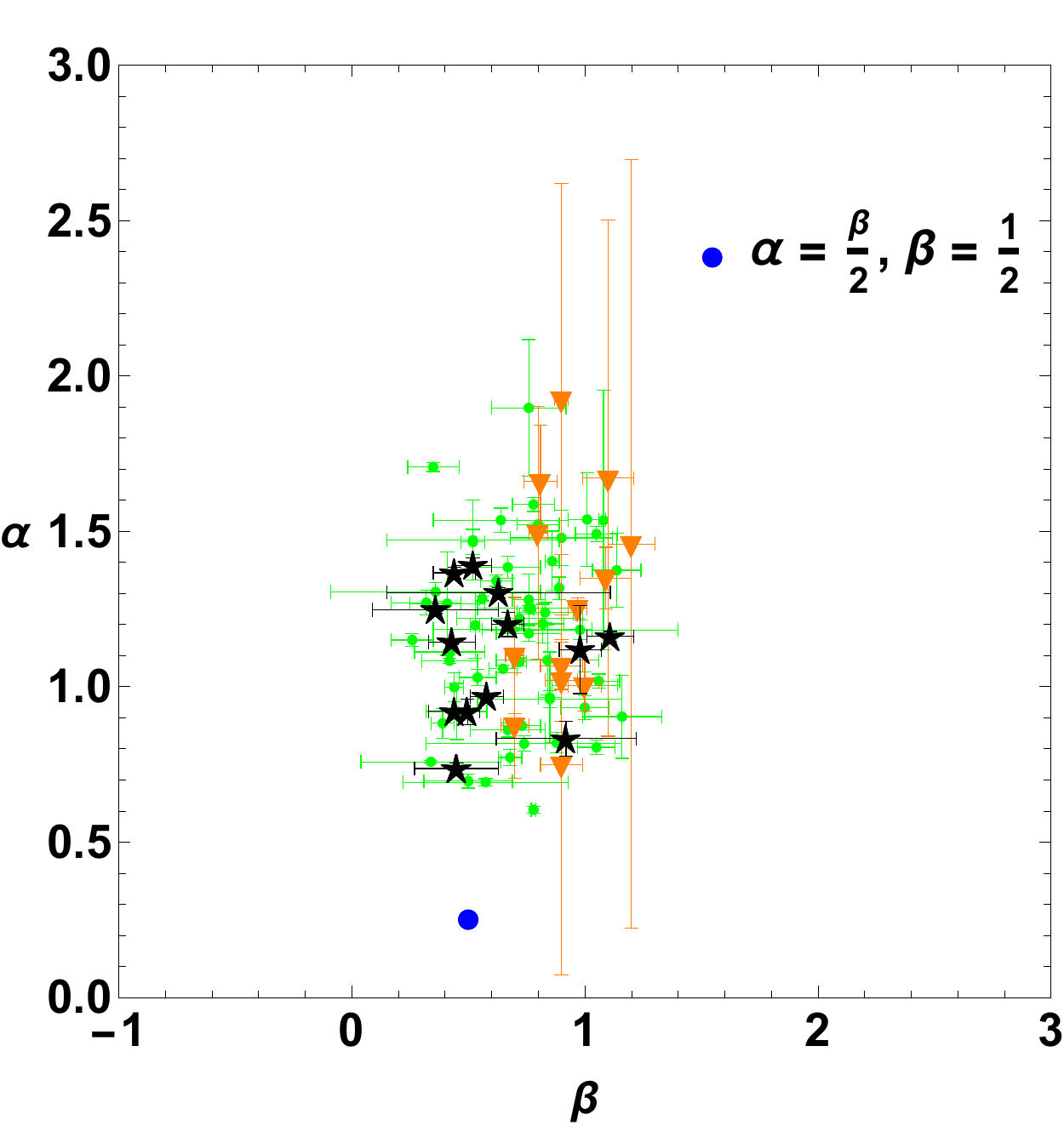}{0.3\textwidth}{k = 0, $v_c < v < v_m$}
\fig{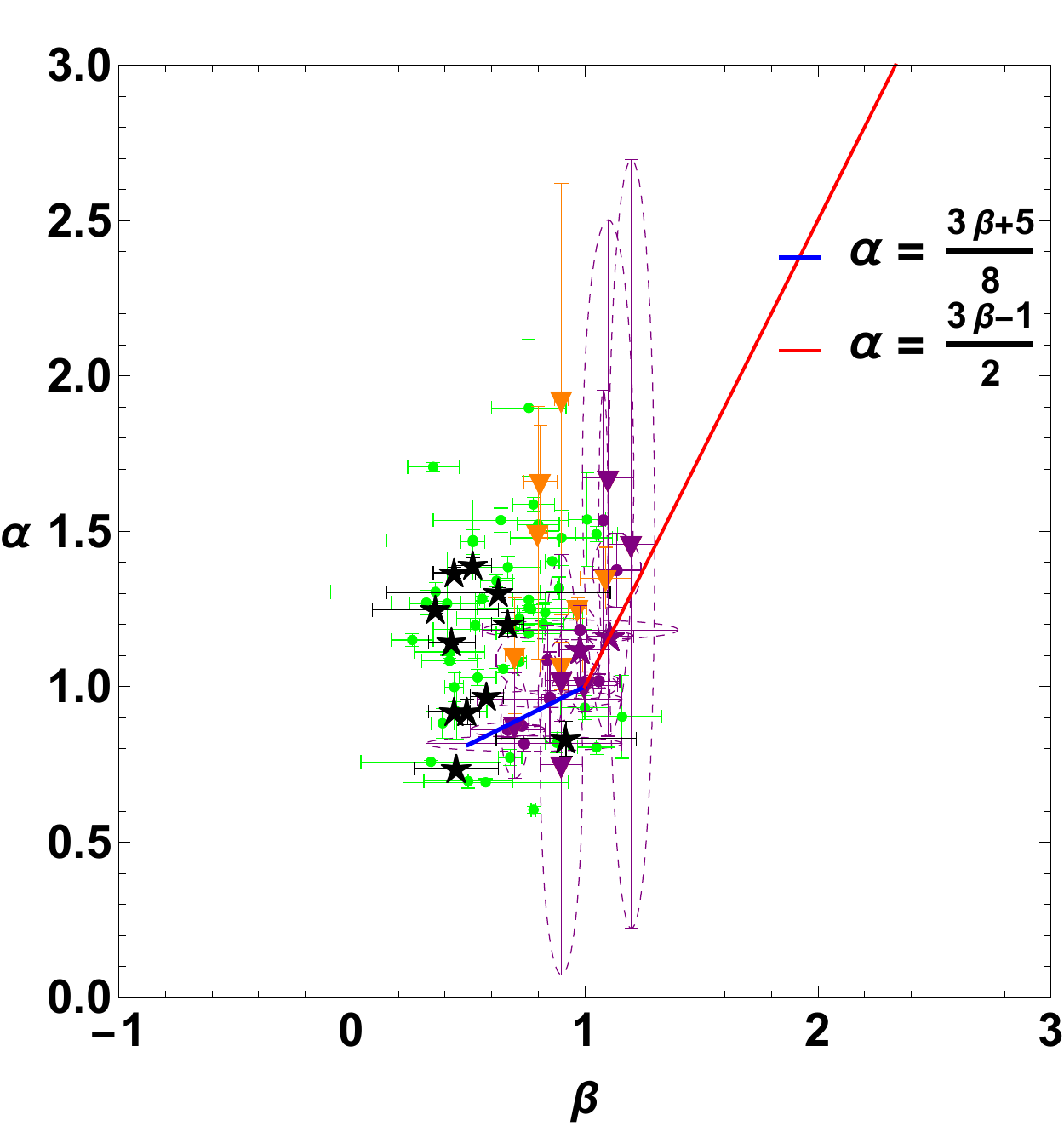}{0.3\textwidth}{k = 0, $v > v_c,v_m$}}
\gridline{\fig{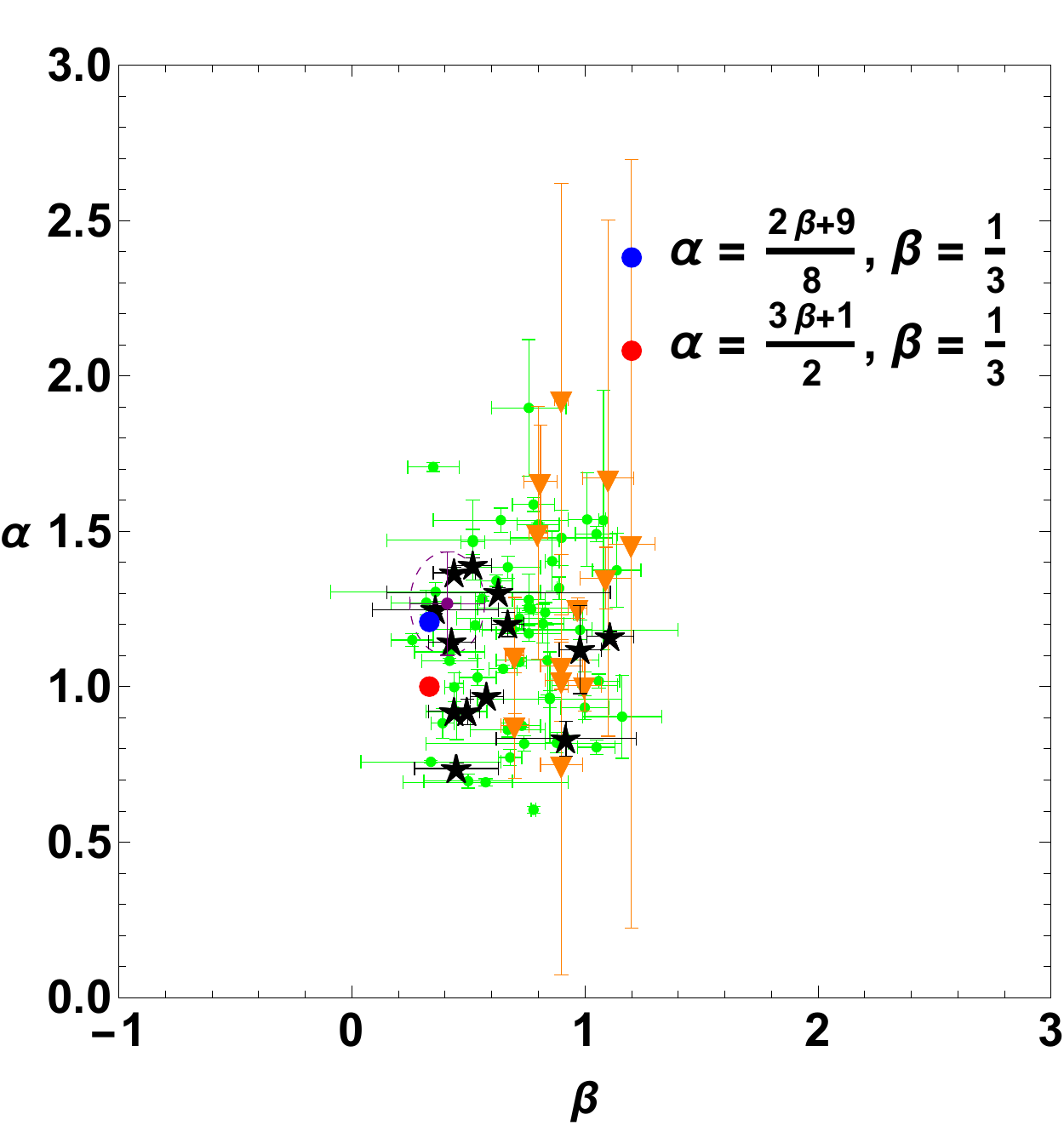}{0.3\textwidth}{k = 2, SC, $v_m < v < v_c$}
\fig{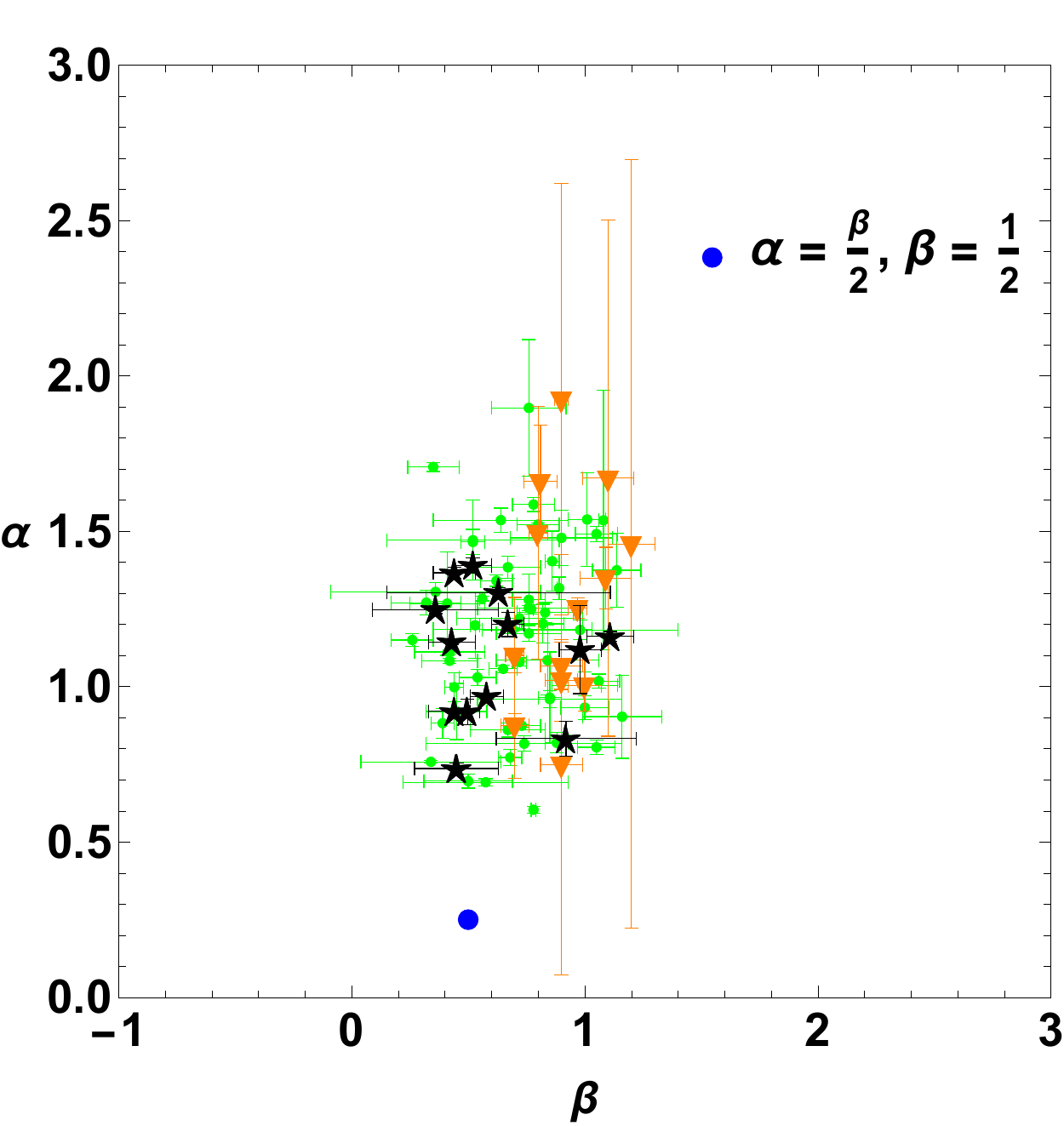}{0.3\textwidth}{k = 2, $v_c < v < v_m$}
\fig{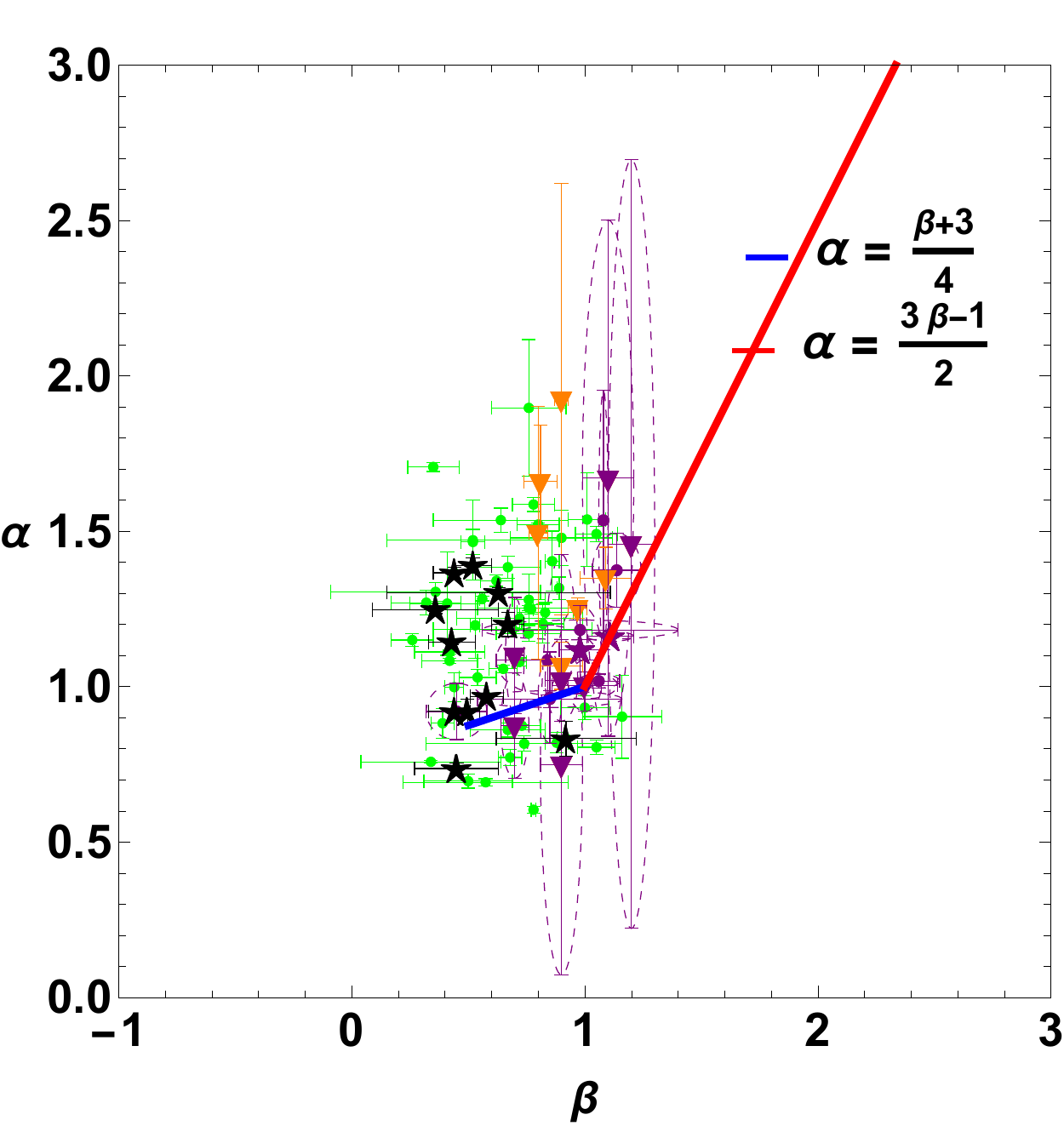}{0.3\textwidth}{k = 2, $v > v_c,v_m$}}

\caption{CRs from the FS model for $k = 0,2$ without energy injection. The relations for $1 < p < 2$ are shown in blue, while the relations for $p > 2$ are shown in red. GRBs that satisfy the relations are shown in purple; SNe-associated GRBs are shown as black stars, GRBs with spectral indices extrapolated from X-ray are shown as orange triangles, other GRBs are shown in green.}
\label{fig:noinj}
\end{figure}

We show the plots of the CRs and fulfilling GRBs in Fig. \ref{fig:noinj}. The CRs are represented by either a line or a point, depending on the CRs of $\alpha$ and $\beta$. CRs (either lines or points) for $1 < p < 2$, are shown in blue, while relation lines or points for $p > 2$ are shown in red. The GRBs which fulfill the CRs are in purple, while those that do not are shown in green. 


It is important to note that the regimes with constant values of $\beta$ are less-preferred over those with varying $\beta$, as those with constant $\beta$ would be represented by a point rather than a line and are therefore less likely to intersect multiple GRBs. Additionally, these single points are not centrally located within the GRB distribution on the $\alpha-\beta$ plane, instead falling at the edge of the observed distribution. This suggests that, even though the single-point CRs are mathematically disadvantaged relative to the other CRs that are extended lines, there is also physical/statistical evidence that they hardly match the GRB set.

For the 13 GRBs-SNe (Table \ref{table:crSNe}), we see that none of the GRBs fulfill either the SC-only or FC-only CRs in either the ISM or Wind environments. The most-preferred regime remains the $\nu > \rm max\{\nu_{\rm c}, \nu_{\rm m}\}$ regime, with the same 2 GRBs fulfilling the CR in both density profiles. We observe no particular clustering of the GRBs-SNe within our data, but the GRBs with extrapolated spectral indices lie at higher values of $\beta$ and have larger error bars.

\begin{table}[ht!]
\caption{Table describing fulfillment of CRs by the sample of 82 optical GRBs. Number of GRBs fulfilling the CR, the cooling regime, the frequency range, the spectral index, the CRs, the proportion compared to the total sample and percentage of the total sample fulfilling each relation are given.}

\begin{center}

\begin{tabular}{ L c C C C C C C c}
 \hline
\multicolumn{9}{c}{No Energy Injection} \\
\hline
n(r) & \text{Cooling} & \nu \text{ Range}    & \beta(p) & \text{CR:}1<p<2      & \text{CR:}p>2 & \text{GRBs} & \text{Proportion} & \text{Figure}\\
 \hline
 r^0    & Slow & \nu_m<\nu<\nu_c & \frac{1}{3} & \frac{6\beta+9}{16} & \frac{3\beta}{2}     & 2 & 2.44 \% & (1a) \\
 
 

r^{-2} & Slow & \nu_m<\nu<\nu_c & \frac{1}{3} & \frac{2\beta+9}{8}  & \frac{3\beta+1}{2} & 1 & 1.22\% & (1c) \\
 

 r^0    & Fast & \nu_c<\nu<\nu_m & \frac{1}{2} & \frac{\beta}{2}  & \frac{\beta}{2}      & 0 & 0\% & (-) \\
 
 
 
 r^{-2} & Fast & \nu_c<\nu<\nu_m & \frac{1}{2} & \frac{\beta}{2}       & \frac{\beta}{2}      & 0 & 0\%  & (-) \\
 

 r^0    & Slow/Fast & \nu> {\rm max \{\nu_c,\nu_m\}} &\frac{p}{2} & \frac{3\beta+5}{8}  & \frac{3\beta-1}{2} & 19 & 23.2\% & (1b) \\
 
 
 
 r^{-2} & Slow/Fast & \nu> {\rm max \{\nu_c,\nu_m\}} & \frac{p}{2} & \frac{\beta+3}{4}   & \frac{3\beta-1}{2} & 18 & 22.0\% & (1d) \\
 
 \hline
\end{tabular}

\end{center}
\label{table:crSummary}
\end{table}




\begin{table}[ht!]
\caption{Table describing fulfillment of CRs by the sample of 13 optical GRBs with SNe associations. Number of GRBs fulfilling the CR, the cooling regime, the frequency range, the spectral index, the equations of the CRs, the proportion compared to the total sample, and the percentage of the total sample fulfilling each relation are given.}
\begin{center}

\begin{tabular}{L c C C C C C C}
 \hline
\multicolumn{8}{c}{No Energy Injection for SNe-associated GRBs} \\
\hline
n(r) & \text{Cooling} & \nu \text{ Range}    & \beta(p) & \text{CR:}1<p<2      & \text{CR:}p>2 & \text{GRBs} & \text{Proportion}\\
 \hline
 r^0    & Slow & \nu_m<\nu<\nu_c & \frac{1}{3} & \frac{6\beta+9}{16} & \frac{3\beta}{2}     & 0 &  0\% \\
 
 

r^{-2} & Slow & \nu_m<\nu<\nu_c & \frac{1}{3} & \frac{2\beta+9}{8}  & \frac{3\beta+1}{2} & 0 & 0\%  \\
 

 r^0    & Fast & \nu_c<\nu<\nu_m & \frac{1}{2} & \frac{\beta}{2}  & \frac{\beta}{2}      & 0 & 0\%  \\
 
 
 
 r^{-2} & Fast & \nu_c<\nu<\nu_m & \frac{1}{2} & \frac{\beta}{2}       & \frac{\beta}{2}      & 0 & 0\%   \\
 

 r^0    & Slow/Fast & \nu> {\rm max \{\nu_c,\nu_m\}} &\frac{p}{2} & \frac{3\beta+5}{8}  & \frac{3\beta-1}{2} & 2 & 15.4 \%  \\
 
 
 
 r^{-2} & Slow/Fast & \nu>{\rm max \{\nu_c,\nu_m\}} & \frac{p}{2} & \frac{\beta+3}{4}   & \frac{3\beta-1}{2} & 2 & 15.4\% \\
 
 \hline
\end{tabular}
\end{center}
\label{table:crSNe}
\end{table}

\subsection{Comparing to CRs in other wavelengths}
\label{sec:compare}

We can compare our results in optical to previous results obtained in X-rays and $\gamma$-rays to better understand the fulfillment of the standard fireball model among different wavelengths. We find 46 GRBs in our optical sample that are coincident with the sample of 222 X-ray GRBs from \citet{Srinivasaragavan2020ApJ}, and 2 GRBs that are coincident with the sample of 86 $\gamma$-ray GRBs from the Second Fermi-LAT GRB Catalog, as analyzed in Dainotti et al. (2022, submitted). We compare the results obtained in optical to the results of analyzing the CRs in X-rays (using Table 1 of \citet{Srinivasaragavan2020ApJ}, taken from \citet{Racusin+09}) and $\gamma$-rays (using Table 3 of Dainotti et al. (2022, submitted), also taken from \citet{Racusin+09}). We summarize the results in Table \ref{table:compare} below.

\begin{table}[ht!]
\caption{Table summarizing fulfillment of CRs in X-ray, optical, and $\gamma$-ray wavelengths for coincident GRBs}

\begin{center}

\begin{tabular}{lccccccc}
 \hline
\multicolumn{2}{l}{} & \multicolumn{3}{c}{ISM} & \multicolumn{3}{c}{Wind}\\
\hline
GRB & Wavelength & SC & FC & $\nu>{\rm max \{\nu_c,\nu_m\}}$ & SC & FC & $\nu>{\rm max \{\nu_c,\nu_m\}}$\\
\hline
050319A	&	O, X	& 	...	& 	...	& 	O, X	& 	...	& 	...	& 	…	\\
050401A	&	O, X	& 	...	& 	...	& 	X	& 	...	& 	...	& 	X	\\
050416A	&	O, X	& 	...	& 	...	& 	...	& 	...	& 	...	& 	…	\\
050730A	&	O, X	& 	...	& 	...	& 	...	& 	...	& 	...	& 	…	\\
050820A	&	O, X	& 	...	& 	...	& 	X	& 	...	& 	...	& 	X	\\
050824A	&	O, X	& 	...	& 	...	& 	X	& 	...	& 	...	& 	X	\\
050922C	&	O, X	& 	...	& 	...	& 	X	& 	...	& 	...	& 	X	\\
060124A	&	O, X	& 	...	& 	...	& 	O	& 	...	& 	...	& 	…	\\
060206A	&	O, X	& 	...	& 	...	& 	X	& 	...	& 	...	& 	X	\\
060526A	&	O, X	& 	...	& 	...	& 	X	& 	...	& 	...	& 	X	\\
060607A	&	O, X	& 	...	& 	...	& 	...	& 	...	& 	...	& 	…	\\
060708A	&	O, X	& 	...	& 	...	& 	X	& 	...	& 	...	& 	X	\\
060714A	&	O, X	& 	...	& 	...	& 	X	& 	...	& 	...	& 	X	\\
060729A	&	O, X	& 	...	& 	...	& 	...	& 	...	& 	...	& 	…	\\
060904B	&	O, X	& 	...	& 	...	& 	O, X	& 	...	& 	...	& 	O, X	\\
060927A	&	O, X	& 	...	& 	...	& 	...	& 	...	& 	...	& 	…	\\
070110A	&	O, X	& 	...	& 	...	& 	O, X	& 	...	& 	...	& 	O, X	\\
070810A	&	O, X	& 	...	& 	...	& 	...	& 	...	& 	...	& 	X	\\
071003A	&	O, X	& 	...	& 	...	& 	X	& 	...	& 	...	& 	…	\\
080310A	&	O, X	& 	...	& 	...	& 	...	& 	...	& 	...	& 	…	\\
080319C	&	O, X	& 	...	& 	...	& 	O 	& 	...	& 	...	& 	O 	\\
080710A	&	O, X	& 	...	& 	...	& 	...	& 	...	& 	...	& 	…	\\
081203A	&	O, X	& 	...	& 	...	& 	O	& 	...	& 	...	& 	O 	\\
090423A	&	O, X	& 	...	& 	...	& 	...	& 	...	& 	...	& 	O 	\\
090426A	&	O, X	& 	...	& 	...	& 	X	& 	...	& 	...	& 	X	\\
090510A	&	O, X, $\gamma$	& 	...	& 	...	& 	O 	& 	...	& 	...	& 	O 	\\
090927A	&	O, X	& 	...	& 	...	& 	X	& 	O	& 	...	& 	X	\\
091020A	&	O, X	& 	...	& 	...	& 	O, X	& 	...	& 	...	& 	O, X	\\
091127A	&	O, X	& 	...	& 	...	& 	...	& 	...	& 	...	& 	…	\\
100418A	&	O, X	& 	...	& 	...	& 	O, X	& 	...	& 	...	& 	O, X	\\
100513A	&	O, X	& 	...	& 	...	& 	O, X	& 	...	& 	...	& 	O, X	\\
100621A	&	O, X	& 	...	& 	...	& 	...	& 	...	& 	...	& 	…	\\
100906A	&	O, X	& 	...	& 	...	& 	O 	& 	...	& 	...	& 	O 	\\
101219B	&	O, X	& 	...	& 	...	& 	...	& 	...	& 	...	& 	…	\\
110503A	&	O, X	& 	...	& 	...	& 	...	& 	...	& 	...	& 	…	\\
110715A	&	O, X	& 	...	& 	...	& 	X	& 	...	& 	...	& 	X	\\
120404A	&	O, X	& 	...	& 	...	& 	...	& 	...	& 	...	& 	…	\\
120711A	&	O, $\gamma$	& 	...	& 	...	& 	...	& 	...	& 	...	& 	…	\\
120907A	&	O, X	& 	...	& 	...	& 	O 	& 	...	& 	...	& 	O 	\\
130606A	&	O, X	& 	...	& 	...	& 	...	& 	...	& 	...	& 	…	\\
140423A	&	O, X	& 	...	& 	...	& 	X	& 	...	& 	...	& 	X	\\
140430A	&	O, X	& 	...	& 	...	& 	X	& 	...	& 	...	& 	…	\\
160227A	&	O, X	& 	...	& 	...	& 	…	& 	...	& 	...	& 	O 	\\
160804A	&	O, X	& 	...	& 	...	& 	O, X	& 	...	& 	...	& 	O, X	\\
161219B	&	O, X	& 	...	& 	...	& 	…	& 	...	& 	...	& 	…	\\
170714A	&	O, X	& 	...	& 	...	& 	…	& 	...	& 	...	& 	…	\\
180205A	&	O, X	& 	...	& 	...	& 	X	& 	...	& 	...	& 	X	\\
\hline
\end{tabular}
\end{center}
\tablecomments{Table summarizes fulfillment of CRs in multiple wavelengths. Column 1 gives the name of the GRB in the optical sample that has also been analyzed in either X-rays or $\gamma$-rays. Column 2 gives the wavelengths for which a particular GRB has been analyzed - O for optical, X for X-rays, and $\gamma$ for $\gamma$-rays. Columns 3, 4, and 5 mark whether a GRB satisfies the given CR in the ISM ($k=0$) environment, while columns 6, 7, and 8 mark whether a particular GRB satisfies the given CR in the Wind ($k=2$) environment. Columns are marked with X, O, or $\gamma$ to represent whether the relation has been satisfied in X-ray, optical, or $\gamma$-ray wavelengths; `...' is placed in all columns for which the GRB does not satisfy the given relation.}
\label{table:compare}

\end{table}

In all wavelengths, we see that none of the coincident GRBs satisfy the fast-cooling, $\nu_c < \nu < \nu_m$ in either the ISM or Wind environments. The SC,  $\nu_m < \nu < \nu_c$ environment is only satisfied by one coincident GRB, 090927A, for the Wind environment in optical wavelengths. Most coincident GRBs satisfy the $\nu>{\rm max \{\nu_c,\nu_m\}}$ regime in either X-ray or optical wavelengths, which is consistent with our expectations. GRBs 060904B, 070110A, 091020A, 100418A, 100513A, and 160804A satisfy this regime for both X-ray and optical in both the ISM and Wind environments, while GRB 050319A satisfies this regime in both X-ray and optical in the ISM environment only. One GRB is in common with our optical sample and the $\gamma$-ray sample in Dainotti et al. (2022, submitted), GRB 120711A, which does not fulfill any CR in either optical or $\gamma$-ray wavelengths.

Only one GRB among our samples, 090510A, is found in all three wavelengths. In X-ray and optical wavelengths, this GRB does not satisfy any of the given CRs. In high energy, we see that GRB 090510A fulfills two relations, corresponding to the $\nu>{\rm max \{\nu_c,\nu_m\}}$ regime and the ISM, SC, $\nu_m < \nu < \nu_c$. The fulfillment of these regimes in high-energy agrees with expectations from previous studies, as the $\nu>{\rm max \{\nu_c,\nu_m\}}$ regime and the SC, $\nu_m < \nu < \nu_c$ are the most commonly fulfilled regions. This analysis, as given in \citet{2021ApJS..255...13D}, has been performed by fitting the LC with a \citet{Willingale2007} model and a time-sliced analysis for the spectral fitting from the time at the end of the plateau and the end time of the observations.
Dainotti et al. (2022, submitted) performed a similar analysis while using a BPL and takes the resulting temporal index and spectral index from the Second Fermi-LAT GRB Catalog. This gives differences in the end-time of the plateau and the spectral index, which ultimately leads to different results.

We note that the distributions of the temporal and spectral indices differ slightly between the optical and X-ray samples. We run the Kolmogorov-Smirnov two-sample test for these distributions, finding a p-value of $p = 0.001$, which indicates that these distributions were not drawn from the same parent sample. However, the KS test has limitations, including increased sensitivity toward the center of the distribution and less toward the tails of the distribution, and sensitivity to differences in multiple characteristics between the two distributions (including location and shape). In this case, the KS test is likely to be too restrictive to provide us with useful results.Thus, we have approached the problem in a different way. Usually the distributions of GRB parameters in X-rays and optical are well fitted by Gaussian distributions (Dainotti et al. 2022 in preparation), thus we perform a less restrictive and more general test by fitting the distributions with a tool in Mathematica called FindDistributionParameters which provides the best fitted distributions. As a result, we obtain that the best fitted distributions are Gaussians.The optical distribution has a mean, $\mu = 0.75$ and a standard deviation, $\sigma = 0.23$, and the X-ray has $\mu = 0.92$ with $\sigma = 0.20$. Thus, the difference between the X-ray and optical samples is not statistically significant - when we fit both distributions with a Gaussian,  Therefore, we find that the distributions are compatible within one $\sigma$.

\subsection{Optical 2D luminosity-time correlation} \label{sec:correlation}

To better understand how the CRs relate to other physical parameters of GRBs, we examine the two-dimensional Dainotti correlation for the sample of optical GRBs which fulfill the most-favored regime in our set of CRs. Following the approach of \citet{delvecchio16}, we investigate the distribution of $\alpha$ values amongst the correlation for the full sample of 82 GRBs to determine if there is any clustering around particular values of  $\alpha$ in the $L_a$- $T^*_a$ relation. We see that the highest value of $\alpha$ corresponds to a log $L_a$ of $\approx 44.5$ erg/s and a log $T^*_a$ of $\approx 4$ seconds, while the lowest value of $\alpha$ corresponds to a log $L_a$ of $\approx 46.5$ erg/s with a log $T^*_a$ of $\approx 2.5$ seconds. We do not observe any particular clustering within this correlation for any values of $L_a$ or $T^*_a$, indicating the temporal index does not have a significant impact on the correlation itself.

This result is different from the one of \citet{delvecchio16}, who examined the distribution of the $\alpha$ parameters along the luminosity-time correlation for a sample of 176 GRB afterglows and found that the $\alpha$ parameters varied systematically with luminosity within the correlation. We do not find this result in our sample, however, we note that \citet{delvecchio16} used a larger sample of GRBs in X-ray wavelengths instead of optical, which may contribute to the observed differences.

We also examine the correlation color-coded by SNe association or X-ray extrapolation in the spectral index to determine if there is any clustering observed. We see that the GRBs-SNe tend to have lower luminosities and larger $T^*_a$ than the other GRBs, located mainly at the lower end of the correlation. This may be because SNe have been observed at lower redshifts than other GRBs that present an optical plateau. There is no observed clustering of GRBs with extrapolated spectral indices, which are scattered within the middle of the distribution.

\begin{figure*}[ht!]
\centering
\includegraphics[width = 0.45\textwidth]{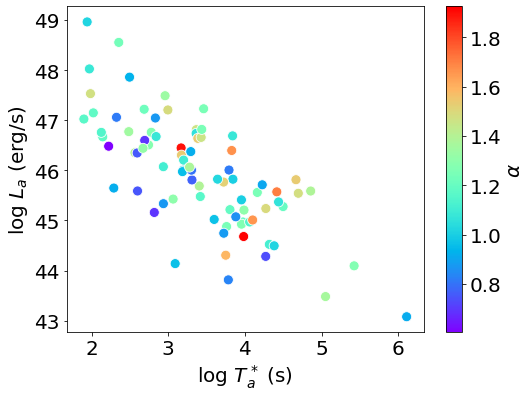}
\includegraphics[width = 0.45\textwidth]{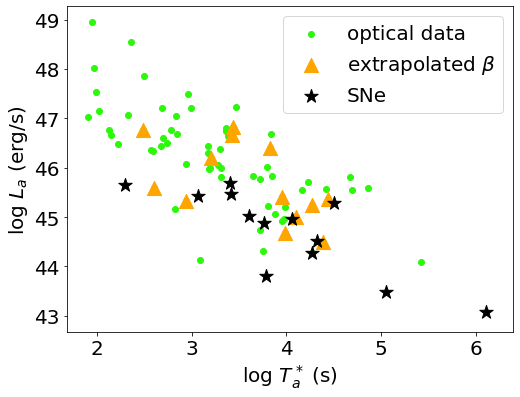}
\caption{Left: Luminosity-time correlation for full sample of 82 GRBs, color-coded by $\alpha$ value for each GRB. Right: luminosity-time correlation for full sample of 82 GRBs, color-coded by class - SNe-associated GRBs are shown as black stars, GRBs with spectral indices extrapolated from X-ray are shown as orange triangles, and all other GRBs are shown as green circles.}
\label{fig:alpha}
\end{figure*}

The correlation of the full sample of 82 GRBs, colored according to the temporal index, is shown in the left panel of Fig. \ref{fig:alpha}. We do not present the same analysis for $\beta$ values as the spectral index is intrinsically correlated with luminosity by definition (see equation \ref{eq:lum}). For completeness of the discussion, the correlation of the full sample color-coded by SNe-association and extrapolated spectral index is shown in the right panel of Fig. \ref{fig:alpha}. 

For the most-favored regime in the set of CRs, we find the 2D luminosity-time correlation for the set of GRBs that satisfies each CR to determine if the scatter can be reduced from that found in \citet{Dainotti2022}. We use the Bayesian D'Agostini method and the {\tt\string cobaya} package in Python to find the best-fit parameters for each of these samples, and analyze the correlation before and after correction for selection bias and redshift evolution. The results are shown in Table \ref{table:corrparams}, which presents the slope of the correlation $a$, the normalization parameter $c$, and the intrinsic scatter $\sigma_{\rm int}^2$. The left columns show the parameters before correction, while the right columns show the parameters after correction (denoted with a superscript $'$). As the BPL model requires four parameters to fit the LCs, we require at least four data points to fit the correlation, thus we only include the regimes and density profiles with $> 4$ GRBs fulfilling the CRs. 

\begin{table}[h!]
\caption{Table of best fit parameters for the 2D luminosity-time correlation for the most-favored regime in each set of CRs} \label{table:corrparams}
\centering
\begin{tabular}{LCCCCCCC}
\hline
\multicolumn{8}{c}{Best fit parameters for no injection, $\nu > \rm max\{\nu_{\rm c}, \nu_{\rm m}\}$  regime} \\ 
\hline
\multicolumn{2}{l}{} & \multicolumn{3}{c}{Uncorrected for Evolution} & \multicolumn{3}{c}{Corrected for Evolution} \\ \hline
\text{Class} & \text{N} & $a$ & $c$ & $\sigma_\text{int}^2$ & $a'$ & $c'$ & $\sigma_\text{int}^{\prime2}$ \\
k=0 & 19 & -1.05 \pm 0.30 & 49.51 \pm 1.03 & 0.92 \pm 0.18 & -0.93 \pm 0.23 & 48.24 \pm 1.00 & 0.75 \pm 0.16 \\
k=2 & 18 & -1.10 \pm 0.28 & 49.16 \pm 1.13 & 0.92 \pm 0.20 & -0.97 \pm 0.24 & 48.36 \pm 1.06 & 0.78 \pm 0.15 \\ 
\hline
\end{tabular}
\end{table}

\begin{figure*}[ht!]
\includegraphics[]{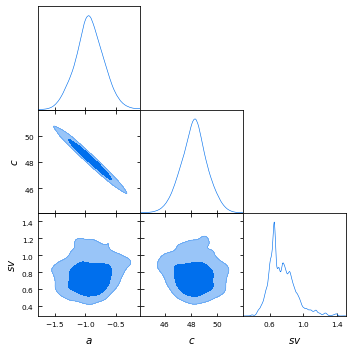}
\caption{Cornerplot showing the distribution of best-fit parameters from the 2D fitting of the ISM environment ($k=0$) after correction, which has the lowest intrinsic scatter. \label{fig:corr}}
\end{figure*}

We find all parameters agree within $1 \sigma$ both before and after correction. We also see that the intrinsic scatter agrees with that of the correlation for the total optical sample from \citet{Dainotti2022} within $1 \sigma$ for both correlations derived from the samples fulfilling both the CRs before correction. After correction, the constant-density ISM ($k=0$) correlation still agrees within $1 \sigma$, while the correlation derived from the sample fulfilling the Wind medium ($k=2$) CR agrees with the intrinsic scatter of the total optical sample within $2 \sigma$. The scatter of the correlations is reduced after correction by 18\% for the ISM and 15\% for the Wind medium, however, we see that on average, the scatter is greater than that found in \citet{Dainotti2022}.
Indeed, $\sigma_{\rm int}^2 = 0.74$ in \citet{Dainotti2022} vs. $\sigma_{\rm int}^2 \approx 0.90$ in our study before correction, and $\sigma_{\rm int}^{\prime2} = 0.57$ in \citet{Dainotti2022}  vs. $\sigma_{\rm int}^{\prime2} \approx 0.80$ in our study after correction. A plot showing the distribution of parameters from the D'Agostini fitting is shown in Fig. \ref{fig:corr} for the correlation built with the 19 GRBs fulfilling the ISM environment in the $\nu>{\rm max \{\nu_c,\nu_m\}}$ regime after correction, which has the smallest intrinsic scatter.


It is important to note that the sample size affects these results, especially when considering the correction for selection bias. The correlations fitted in \citet{Dainotti2022} were found using the entire sample of 99 GRBs, whereas the results here are found using a maximum of 19 GRBs. For the correlations that have been corrected for evolutionary effects, the large scatter is, again, largely due to the small sample size - the Efron \& Petrosian method requires a large sample size to work effectively, so the correction applied to the correlations here is based on the larger optical sample given in \citet{Dainotti2022}. We also note that the 3D correlation between $L_a$, $T^*_a$, and the peak luminosity $L_{\rm peak}$ considered in \citet{Dainotti2022} is not studied here as the sub-sample of GRBs with observed $L_{\rm peak}$ is too small to test additional results.

\section{Discussion and Conclusion} \label{sec:conclusion}

We have tested a complete set of CRs, taken from \citet{2019ApJ...883..134T}, \citet{Racusin+09}, and Dainotti et al. (2022, submitted), on a set of 82 GRBs observed in optical wavelengths and fitted with a BPL. We see that only 24/82 GRBs fulfill a CR in our sample, suggesting that our sample does not fulfill the expectation of the standard fireball model. The most favored environment for our set of CRs is the $\nu > \rm max\{\nu_{\rm c}, \nu_{\rm m}\}$ regime, with 23.2\% fulfillment for $k=0$. The $k=2$ CR in this regime shows 22\% fulfillment, indicating high preference (comparatively) for this regime, but not favoring either the ISM or Wind environments. We see no fulfillment for the FC-only regime for either environment, indicating a very low preference for this regime. This lack of preference may occur because the FC-only relations have constant $\beta$, which means their CRs are points, rather than lines as in the $\nu > \rm max\{\nu_{\rm c}, \nu_{\rm m}\}$ regime.

We now compare our results with previous results in the literature. \citet{oates2012} examined a sample of 48 optical LCs observed by \emph{Swift} and tested whether a correlation between the luminosity and the optical decay index after 200 seconds could be explained by the standard model. They tested three CRs in frequency ranges without injection - $\nu_{\rm c} < \nu_{\rm opt}$, $\nu_{\rm m} < \nu_{\rm opt} < \nu_{\rm c}$ in a Wind medium, and $\nu_{\rm m} < \nu_{\rm opt} < \nu_{\rm c}$ in a constant-density medium, of which we consider both $\nu_{\rm m} < \nu_{\rm opt} < \nu_{\rm c}$ regimes. They found no particular clustering of the $\alpha$ and $\beta$ parameters around a specific CR, indicating that the afterglows of those optical GRBs are not well-explained by the basic standard model. Our results agree with this study, as many of the GRBs in our sample do not agree with the tested CRs, which would indicate the standard model is not the optimal explanation for these optical GRBs. Regarding specific CRs, the authors found $\approx 6$ GRBs satisfying the $\nu_{\rm m} < \nu_{\rm opt} < \nu_{\rm c}$, $k=0$ CR, where we find 2 GRBs, and $\approx 4$ GRBs satisfying the $\nu_{\rm m} < \nu_{\rm opt} < \nu_{\rm c}$, $k=2$ CR, where we find 1 GRB. The difference in these results may be because \citet{oates2012} used the slope of the LC after 200 seconds, while we use the slope after the end-time of the plateau emission, and the sample size and GRBs are different. However, regardless of these differences, we agree that a more complex model of the afterglow, including the modeling of continuous energy injection, could give a better explanation of the optical afterglow and this sample of GRBs.


Continuing on the comparison with other studies, but in different wavelengths, we find some discrepancies with studies of CRs in higher energies, namely $\gamma$-rays and X-rays. 
In high-energy, \citet{2019ApJ...883..134T} studied a sample of 59 GRBs detected by Fermi-LAT. They found that a high proportion of their sample satisfies the CRs, indicating that it can be well-explained by the standard model. They found their most-preferred regimes to be the SC, $\nu_{\rm m} < \nu < \nu_{\rm c}$ regime for the ISM/wind media and the $\nu > \rm max\{\nu_{\rm c}, \nu_{\rm m}\}$ regime for the ISM. We see in our results that the $\nu > \rm max\{\nu_{\rm c}, \nu_{\rm m}\}$ is similarly preferred for the sample, though we see no strong preference for the ISM compared to the Wind medium. We also see a similar preference for SC-only regimes over FC-only regimes, though the rates of satisfaction for SC-only regimes are lower in our optical sample than in the high-energy sample. Although in our study we are not able to exclude $k=2$ using only the CRs, physical arguments \citep{2002ApJ...568..820G} indicate that optical observations of GRBs are unlikely to satisfy $\nu>\nu_{\rm c}$ with $k=2$, especially in the SC regime. Even for $k=0$, the assumption that $\nu > \max\{\nu_c,\nu_m\}$ places restrictions on the parameters of the GRB. 

Again in high energy, \citet{2021ApJS..255...13D} studied three GRBs - 090510, 090902B, and 160509A - and checked the fulfillment of the CRs without energy injection. The results show that an SC environment is preferred for all three GRBs, with an ISM environment preferred for GRB 090510 and a Wind environment preferred for GRB 090902B and GRB 160509A. 

In X-rays, \citet{Racusin+09} studied 230 GRB afterglows detected by \emph{Swift} for a set of CRs with and without energy injection in a constant-density ISM or Wind medium. In general, they found that the ISM is preferred over the Wind environment, which agrees with our results for most cases - the ISM is very slightly preferred over Wind for the SC-only regime, and in the $\nu > \rm max\{\nu_{\rm c}, \nu_{\rm m}\}$ regime. However, for all other relations, we see no clear preference for one over the other. 

\citet{2017ApJ...844...92F} used numerical models of X-ray and optical LCs, testing CRs in SC and FC regimes within an ISM environment. They applied their findings to the X-ray and optical afterglow of GRB 130427A and saw that discrepancy in the temporal decay indices between the wavelengths implies that the standard ISM model assuming a single emission component is not enough to describe the emission of this GRB.\\

It is worth noting that in the case of ISM, the spectral breaks evolve as $\nu_m\propto t^{-\frac32}$ and $\nu_c\propto t^{-\frac12}$. Therefore, for late-time observations (e.g. where $t_{\rm d}$, the deceleration time is $\approx 10^4$ seconds and in this case we assume $T_a=t_d$) the optical frequency is expected to be in the cooling regime $\nu_{\rm opt}> {\rm max \{\nu_c, \nu_m\}}$. In this case, the magnetic microphysical parameter (${\rm \epsilon_{B}}$) would be constrained by the following equation \citep[Eq. 11 in][for the adiabatic regime]{Sari+98}:

\begin{equation}\label{5}
    \rm \epsilon_{B}> 1.03\times 10^{-2} \left(\frac{1+z}{1.54}\right)^{-\frac13} n^{-\frac23}   \left(\frac{\eta}{0.2}\right)^\frac13\,  E_{\rm iso,51.8}^{-\frac13}\, t^{-\frac13}_{d}\, \nu_{\rm 1\,eV}^{-\frac23},
\end{equation}

where $E_{\rm iso}$ is the isotropic radiated energy, $z$ is the redshift, $\eta$ is the kinetic efficiency to convert kinetic energy to radiation, and $n$ is the density medium.  Thus, if we consider as an example GRB 060729 \citep{2010ApJ...711.1008G}, with $z = 0.54$, $t_{\rm d} = 75857.8=10^{4.88}$ s, $E_{\rm iso}=6.7 \times 10^{51} {\rm erg}$, assuming the constant density medium of 1 particle/${\rm cm^{3}}$, and an efficiency of $\eta=20\%$, then we have $\epsilon_B> 1.03 \times 10^{-2}$. This value is consistent with the derived ones in the afterglow synchrotron scenarios \citep[e.g., see][]{2014ApJ...785...29S}.

\citet{Srinivasaragavan2020ApJ} also conducted a study on X-ray afterglows and again found a significant percentage of their sample fulfilled their relations, which disagrees with our results. However, they found that the SC-only regimes in either a Wind ($k=2$) or ISM ($k=0$) environment were most favored and the FC-only regimes were disfavored, which agrees with our study.


\citet{wang15} studied a sample of 85 GRBs (as compared to our sample of 82 with only 34 GRBs in common with their sample) with X-rays and optical afterglows and found they could divide their sample into three categories based on their agreement with the CRs: 45 GRBs are compatible with the fireball model in all segments of the afterglow, 37 GRBs are compatible with the fireball model in at least one segment of the afterglow, and 3 are incompatible with the fireball model. 

Comparing our results to CRs in lower energy, we conclude that our results largely agree with studies of CRs in radio as in \cite{2021ApJ...911...14K,2021MNRAS.504.5685M}.  
Similar to \citet{oates2012}, both studies suggest a more complex model of GRB afterglows may be needed to accurately describe the lower-energy emission in GRB afterglows. 

We again emphasize that optical afterglows present additional complications compared to high-energy emission or X-ray and radio afterglows, so the incompatibility of our results with the standard fireball model and the lack of agreement with studies in high-energy or X-ray wavelengths is not unexpected. 

Regarding the search for a standard set of GRBs that could be driven by a given environment or regime, we analyze the two-dimensional luminosity-time correlation for the sample of GRBs that fulfill the most-preferred CRs, both before and after correction for evolutionary effects. We find that the slope of the correlation agrees with $-1$ within $1 \sigma$ for both $k$, both before and after correction, which is in line with our expectations from previous studies in X-ray \citep{Dainotti2013b, Srinivasaragavan2020ApJ,2021PASJ...73..970D}, optical \citep{Dainotti2020ApJb, Dainotti2022}, and radio \citep{2022ApJ...925...15L} wavelengths. However, we also note that the small sample sizes inhibit our ability to accurately compute the intrinsic scatter of the correlation, especially after correction. We additionally were not able to compute the scatter of the 3D fundamental plane in optical wavelengths for these GRBs due, again, to the small sample size.

We compare our results with \citet{Srinivasaragavan2020ApJ}. They considered the 3D fundamental plane relation in X-rays to test if the GRBs that fulfilled particular CRs could be used to reduce the scatter of the 3D plane relation, similar to the analysis used in this study. They find that GRBs in all categories of CRs have scatters that agree with that of the ``Gold'' sample in X-rays within $1 \sigma$, except for the FC-only categories. For the FC-only regimes, in both ISM and Wind environments, they find that the scatter is reduced from the ``Gold'' sample and other samples found in the literature. Due to the paucity of our sample compared to previous studies, we analyze the 2D correlation rather than the 3D fundamental plane and compare it to the total sample rather than the Gold sample. We see that generally the scatter of each correlation agrees with the scatter of the 2D correlation for the total sample within $1 \sigma$ both before and after correction, though none of the CR groups reduce the scatter even after correction for selection bias. However, we again note that the lack of regimes with reduced scatter, especially after correction, is likely due to the small sample sizes obtained in each regime. 

In conclusion, we find: 
\begin{enumerate}
    \item Though expected, the majority of our sample does not satisfy the CRs, indicating that a more complex explanation may be needed to accurately model GRB optical afterglows.
    \item For those CRs that are satisfied, we find a preference for the $\nu > {\rm max \{\nu_{\rm c}, \nu_{\rm m}\}}$ regime. The FC-only regime is the least preferred. However, we here note that the $\nu > {\rm max \{\nu_{\rm c}, \nu_{\rm m}\}}$ ordering is unlikely for optical wavelengths when $k=2$, as it would  put severe restrictions on the GRB's parameters \citep{2002ApJ...568..820G}.
    \item For the most-preferred relations, in the majority of cases, the intrinsic scatter of the luminosity-time correlation agrees with that of the total correlation (given in \citet{Dainotti2022}) within $1 \sigma$ both before and after correction for selection bias and redshift evolution.
    \item The single-point CRs
    lie at the edge of the $\alpha$-$\beta$ distribution. In addition, they are mathematically disadvantaged, because they are singular points. This is additional evidence that the standard FS scenario in this simplified version may not be the optimal one to interpret the optical emission and more complex modeling is required.
   \item The comparison among optical, X-rays and high-energy $\gamma$-rays for the GRBs that are in common with the sample analyzed here shows that the fulfillment of the $\nu>{\rm max \{\nu_c,\nu_m\}}$ and the SC, $\nu_m < \nu < \nu_c$ regimes in X-rays and $\gamma$-rays is the highest among all CRs. This agrees with the expectations from previous studies, as in high energy these are the most commonly fulfilled regions.
\end{enumerate}

Eventually, with a larger sample of observed optical GRBs, we can reduce the uncertainty in our results and better understand the implications of these CRs with the optical data.

\section*{Acknowledgements}
We would like to thank Bing Zhang for his comments while writing this paper. DL acknowledges the support of the U.S. Department of Energy, Office of Science, and Office of Workforce Development for Teachers and Scientists (WDTS) under the Science Undergraduate Laboratory Internships (SULI) program. We thank Aleksander Lenart for his initial work on the CR Notebook. NF acknowledges financial support from UNAM-DGAPA-PAPIIT  through grant IN106521. We would also like to acknowledge the support of the NAOJ Division of Science in making this research possible.





\bibliographystyle{mnras}
\bibliography{opticalCR}




\end{document}